\documentclass{pj}

\RequirePackage{manfnt}

\RequirePackage{amsthm}
\newcommand{\bm}{\mathbfit}

\usepackage{booktabs}
\RequirePackage{bm}
\RequirePackage{siunitx} % Angabe von Einheiten

\RequirePackage{color} 

\RequirePackage{amssymb}  
\RequirePackage{amsmath}
\RequirePackage{cleveref}
\RequirePackage{relsize}

\RequirePackage{upgreek}
\RequirePackage{bm}

\RequirePackage{esvect}

\let\originalleft\left
\let\originalright\right
\def\left#1{\mathopen{}\originalleft#1}
\def\right#1{\originalright#1\mathclose{}}

\newcommand{\rem}[1]{}
\newcommand{\veg}[1]{\bm{#1}}     % geometrical / physical vectors
\newcommand{\dd}{\mathrm{d}}  % The differential symbol is actually an operator

\newcommand{\jm}{\mathrm{j}}  % For engineer stuff 

\newcommand{\e}{\mathrm{e}}

\RequirePackage{mathtools}

\DeclareMathOperator{\ee}{e}

\newcommand\restr[2]{{% we make the whole thing an ordinary symbol
		{\left.\kern-\nulldelimiterspace % automatically resize the bar with \right
		#1 % the function
		\vphantom{|} % pretend it's a little taller at normal size
		\right|}_{#2} % this is the delimiter
}}

\newcommand\rst[3]{{% we make the whole thing an ordinary symbol
		\left.\kern-\nulldelimiterspace % automatically resize the bar with \right
		#1 % the function
		\vphantom{|} % pretend it's a little taller at normal size
		\right|_{#2}^{#3} % this is the delimiter
}}

\usepackage{mathrsfs}

\usepackage{graphicx}
\usepackage{amsmath}
\usepackage{amssymb}
\usepackage{bm}
\usepackage{upgreek}
\usepackage{multirow}
\usepackage{overpic}
\usepackage{esint}

\def\ar{\begin{array}}
 \def\arr{\end{array}}
\def\be{\begin{eqnarray}}
 \def\en{\end{eqnarray}}
\def\bee{\begin{equation}}
 \def\ee{\end{equation}}

 \usepackage{etoolbox}

\title{An Indoor Localization Technique Utilizing Passive Tags and 3-D Microwave Passive Radar Imaging}
\author[1,\,*]{Quanfeng~Wang}
\author[1]{Alexander~H.~Paulus}
\author[2]{Mei~Song~Tong}
\author[1]{Thomas~F.~Eibert}

\shortauthor{LastName1 et\,al.}

\shortauthor{Wang et\,al.}

\affil[1]{Department of Electrical Engineering, School of Computation, Information and Technology, Technical University of Munich, Munich, Germany}
\affil[2]{Department of Electronic Science and Technology, Tongji University, Shanghai, China}
\emails{Corresponding authors:~Quanfeng Wang (quanfeng.wang@tum.de).}

\StartPage{1}
\volume{xxxx}{xx}{xxx}{xx}
\paperdoi{10.2528/PIER24120903} % DOI
\history{Date Month Year}{}{xxx}{xxx}

\AddToHook{shipout/foreground}{
  \put(0.5\paperwidth, -0.8cm){
    \makebox[0pt][c]{
      \begin{minipage}{\textwidth}
        \centering \scriptsize
        This paper is published in Progress In Electromagnetics Research (PIER), Vol.181, pp.89--98, 2024. This is the author's version which has not been fully edited and content may change prior to final publication. This repository copy is  provided to comply with open-access requirements.
      \end{minipage}
    }
  }
}

\begin{document}

\begin{Abstract}
	A privacy-compliant indoor localization approach utilizing a 3-D near-field (NF) passive radar imaging technique is presented. This technique leverages ubiquitously radiated electromagnetic fields for imaging, with passive tags introduced to enhance the strength of scattering fields, thereby enabling precise localization at the imaging level. The method also supports localization in non-ideal imaging scenarios, such as for limited bandwidth or in highly-reflective environments. Based on their geometrical properties the simple and low-cost passive tags enable intuitive differentiation between individuals or objects. Associated privacy protection mechanisms are discussed, where the frequency-varying properties of the passive tags provide additional flexibility and potential applications under privacy and ethical considerations. Several forms of passive tags are presented, where both simulation and experimental results validate the effectiveness of the proposed passive tag designs.
\end{Abstract}

% \begin{multicols}{2}
\section{Introduction}
\lettrine{\color{titlecolor}{O}}{utdoor}
localization techniques and their applications have become relatively mature over the past decades, while indoor localization technologies have emerged as a popular research area more recently and are currently extensively investigated~\cite{alam2021devicefree}. Indoor localization involves estimating the position of individuals or objects within enclosed environments, which is critical for applications in sectors such as healthcare~\cite{arcadiustokognon2017structural}, disaster management~\cite{zelenkauskaite2012interconnectedness}, smart buildings~\cite{snoonian2003smart}, the Internet of Things (IoT)~\cite{zanella2014internet}, and machine type communication (MTC)~\cite{taleb2012machine}, where precise location data enhances services, security, and operational efficiency.

With the advancement of wireless communication technologies and the widespread use of portable electronic devices, radio frequency (RF)-based indoor localization has gained considerable attention~\cite{zafari2019survey}. Examples include indoor localization systems based on wireless communication technologies such as WiFi~\cite{feng2012receivedsignalstrengthbased,paul2009rssibased,shan1985spatial,wang2017csibased,luo2017pallas}, Bluetooth~\cite{diaz2010bluepass,gonzalez-castano2002bluetooth,zafari2016microlocation}, and ZigBee~\cite{baronti2007wireless}. At the same time, diverse localization algorithms, which utilize information such as received signal strength indicator (RSSI)~\cite{feng2012receivedsignalstrengthbased}, channel state information (CSI)~\cite{wang2017csibased}, angle of arrival (AoA)~\cite{kumar2014accurate}, time of flight (ToF)~\cite{xiong2015tonetrack}, time difference of arrival (TDoF)~\cite{xu2006position}, and fingerprinting~\cite{wang2017csibased}, have been successfully applied in this field. 

Utilizing ubiquitous radiation sources such as those from WiFi without dedicated devices, is essentially the idea of passive radar. However, this approach presents significant challenges. For instance, the RSSI-based method suffers from very poor localization accuracy, particularly in through-the-wall scenarios and environments with multipath fading~\cite{zafari2019survey,yang2013rssi}. Similarly, the AoA-based technique is highly sensitive to inaccuracies in the estimated AoA~\cite{kumar2014accurate}. ToF- and TDoF-based techniques, on the other hand, demand strict synchronization between transmit (Tx) and receive (Rx) devices and a high time-domain sampling rate~\cite{zafari2019survey}. Fingerprinting techniques estimate the location of targets of interest (TOIs) by comparing received signals with a pre-recorded signal strength map of the entire target environment. However, localization results can be unreliable in realistic scenarios~\cite{maheepala2020lightbased}.

% From the perspective of utilized sources, the proposed technique is based on passive radar, which aims to utilize ubiquitous radiation sources, such as the communication systems like WiFi. However, from an algorithmic perspective, the proposed method employs microwave imaging, which differs from many existing techniques, such as fingerprinting-based WiFi localization~\cite{wang2017csibased}. Fingerprinting techniques estimate the location of targets of interest (TOIs) by comparing received signals with a pre-recorded signal strength map of the entire target environment. However, localization results can be unreliable in realistic scenarios~\cite{maheepala2020lightbased}. In contrast, the imaging technique generates holographic images of the TOIs from electromagnetic fields at sufficient enough measurement positions, enabling precise localization and providing additional visual information.

Particularly in RF-based indoor localization, RF identification device (RFID) technology is widely used~\cite{holm2009hybrid,ni2003landmarc,shirehjini2012rfidbased,mariotti2012wireless,willis2004passive}. It is based on the transmission of data between RFID tags and corresponding RFID readers, where sub-meter accuracy is, however, hard to achieve. Active RFID tags work with a specific power supply, enabling communication over several hundred meters with low-cost circuit design, but they are commonly not integrated into portable user devices~\cite{zafari2019survey}. On the other hand, passive RFID tags operate without batteries, as they extract the required energy from the illuminating electromagnetic waves and their communication range is, thus, considerably shorter~\cite{zafari2019survey}.

Besides RF-based indoor localization, vision-based indoor localization systems, also known as camera-based or image-based localization techniques, have been applied in areas such as surveillance~\cite{remagnino2002videobased} and crowd counting~\cite{hou2011people}. These systems typically rely on specific imaging devices, e.g., Kinect cameras are used under good visible light conditions~\cite{xia2011human}, while infrared cameras are employed for thermal imaging in low-visibility environments~\cite{lu2016wherea}. Building on this premise, previous research has primarily focused on extracting precise location information from the corresponding images. Many related techniques, such as color histograms~\cite{ravi2007indoor}, automatic indexing methods~\cite{wannous2012place}, and transfer learning~\cite{kawaji2010imagebaseda,lu2016wherea} have been proposed. 

One of the most criticized aspects of these vision-based methods is the use of cameras, which raises significant concerns regarding user privacy~\cite{rapoport2012home}. In addition to the fact that location information is inherently sensitive for users, the use of optical or infrared cameras for indoor scene imaging can cause significant discomfort. In fact, most existing indoor localization systems have not adequately addressed privacy and ethical concerns~\cite{zafari2019survey}, as they primarily focus on achieving high localization accuracy. This undoubtedly hampers the development and commercialization of localization technologies.

In this work, we discuss an indoor localization technique based on near-field (NF) passive radar microwave imaging that utilizes passive tags for accurate detection and identification of TOIs. This method can be regarded as a hybrid approach combining RF-based and vision-based indoor localization. The technique relies on the concept of passive radar, adapts the idea of tags known from the context of RFID technology, and is able to provide images of the indoor scenery, similar to the information generated by camera-based systems. In contrast to active microwave imaging, the presented passive technique avoids active scanning, such as the movement of Tx/Rx pairs in monostatic synthetic aperture radar (SAR) or cooperative scanning systems in multiple-input-multiple-output SAR imaging setups. Since no additional radiation is emitted, the presented approach reduces potential interference with and disturbance of existing communication systems in the indoor environment. In this context, NF refers to complex electromagnetic environments such as indoor scenarios, in contrast to far-field cases commonly encountered in remote sensing. With the assistance of passive tags, the position of TOIs can be accurately reconstructed by a holographic passive imaging technique, even under conditions of limited imaging quality and resolution. 

In this context, it is worth mentioning that the terms passive and active in localization are often defined such that active localization determines the location of specific targets by localization devices or tags carried by the TOIs, whereas passive localization does not rely on any devices or tags, nor does it require any participation from the TOI during the localization process~\cite{alam2021devicefree}. Based on this classification, the localization technology discussed in this work technically falls under the category of active localization. However, the employed tags differ significantly from most existing techniques, especially RFID, as they are completely passive, i.e., they do not require batteries and they do not need to extract energy from the incident fields, which relieves the range limitations encountered with passive RFID tags. In addition, various geometrical properties can be easily incorporated for identification purposes. Lastly, the non-optical nature of the microwave images drastically reduces ethical concerns, while operation of the system does not require a visually bright illumination, which is commonly associated with and required by optical cameras.

In Section~\ref{sec:method}, the localization configuration with passive tags and the imaging algorithms are briefly introduced. Numerical and experimental results are presented in Section~\ref{sec:numRes} and Section~\ref{sec:meaRes}, respectively. Privacy and ethical considerations related to the passive tag design are briefly provided. Finally, conclusions are drawn in Section~\ref{sec:conclusion}.

\section{Configuration and Algorithm}
\label{sec:method}
\begin{figure}[t]
	\centerline{\includegraphics[width=0.95\columnwidth,draft=false]{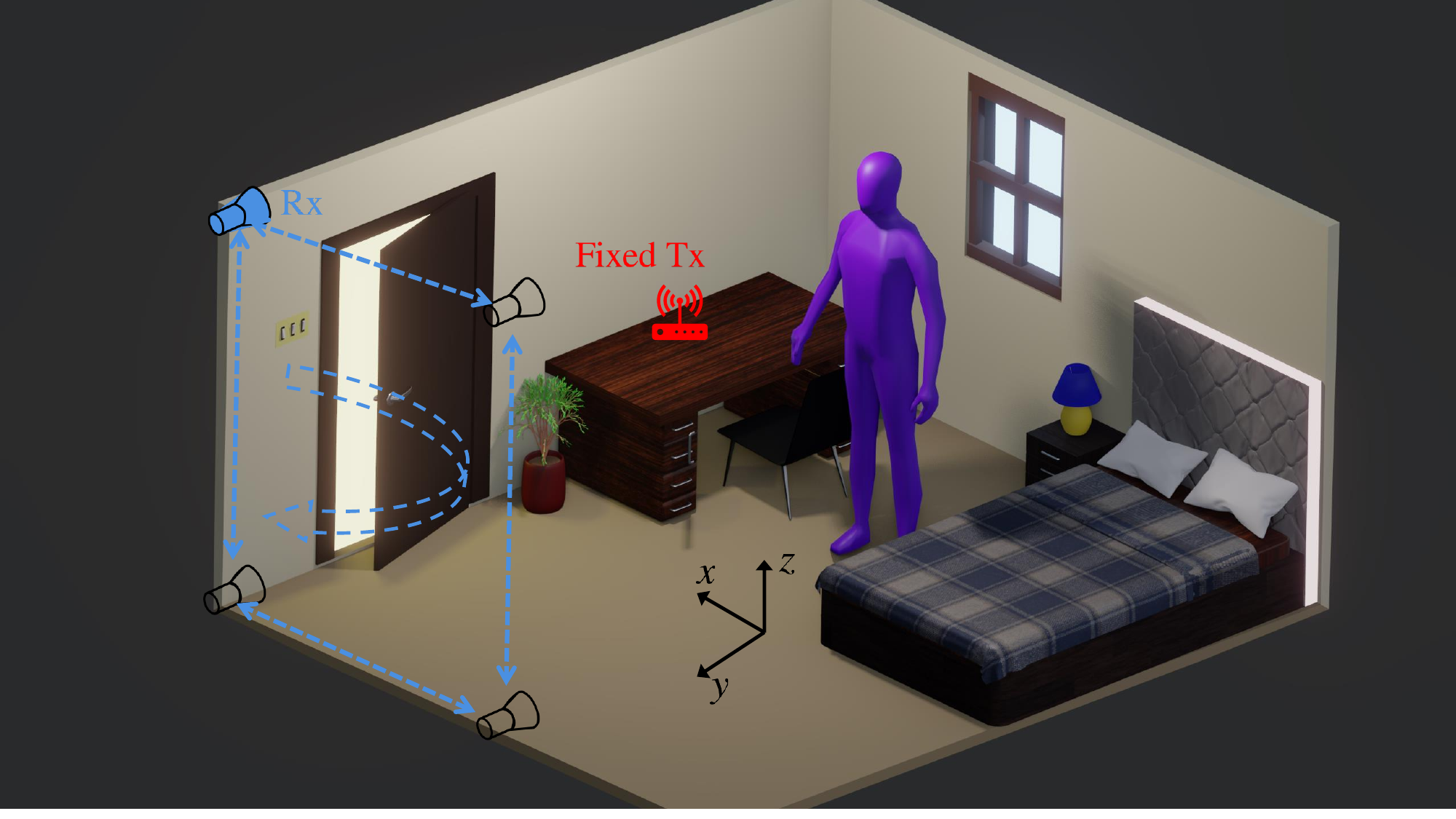}}
	\caption{Typical configuration for indoor localization in a passive radar setup within a bedroom, where a fixed Tx serves as the illumination source. The observation signals across the observation area are collected by a single scanning Rx antenna. The TOIs to be localized may include persons or objects within the room.
	 }
	\label{fig:config}
\end{figure}

The configuration under consideration based on a passive radar setup is illustrated in Figure~\ref{fig:config}, where a single static Tx antenna serves as the illumination source. Observation field samples are collected by either a single scanning Rx antenna moving across the entire sampling area, or more preferably, by a receiving antenna array. A step size of half a wavelength $\lambda/2$ may be used for sampling. All contributions including incident fields $\veg{E}_{\mathrm{i}}$ directly from the Tx antenna, scattering fields $\veg{E}_{\mathrm{s}}$ from the TOIs, and all other parasitic echo scattering fields $\veg{E}_{\mathrm{sp}}$ are captured by the Rx antenna.

Passive tags are attached to the TOIs to enhance the strength of the scattering fields and serve as beacons for localization. The receiving signals from the Rx antenna are then post-processed by the imaging algorithm in~\cite{wang2024TAP}. The process is briefly summarized here. The output signal from the probe at observation position $\veg{r}_{m}$ is expressed as (a time dependency of $\e^{\,\jm \omega t}$ is assumed and suppressed)~\cite{eibert2015electromagnetic,wang2024TAP}
\begin{align}
    U(\veg{r}_{m})=&\iiint_{V_{\rm w}}\veg{w}(\veg{r}-\veg{r}_{m}) \cdot  \notag\\
	&\bigg[\veg{E}_{\rm i}\left({\veg r},\veg{J}_{\rm i}\right)
   +\veg{E}_{\rm s}\left({\veg r},\veg{J}_{\rm s}\right) + \veg{E}_{\rm sp}\left({\veg r},\veg{J}_{\rm sp}\right)\bigg] \dd v\,,\label{eq:Urm}
\end{align}
where $\veg{w}(\veg{r}-\veg{r}_{m})$ is a vector weighting function describing the receiving behavior of the Rx antenna with volume $V_{\rm w}$. By introducing a propagating plane-wave representation of the underlying Green's function, \eqref{eq:Urm} can be written in terms of the plane-wave spectra (PWS) of equivalent sources representing the Tx antenna ($\tilde{\veg{J}}_{\rm i}$), the TOIs ($\tilde{\veg{J}}_{\rm s}$), and parasitic echoes ($\tilde{\veg{J}}_{\rm sp}$). The PWS are expanded by directive vector spherical harmonics $\bm{D}^{(1/2)}$ according to ~\cite{ostrzyharczik2023inverse}
\begin{equation}
	\tilde{\veg{J}}_{\rm i/s/sp}(\veg{k},\veg{r}'_{\rm i/s/sp})=\sum\limits_{n=1}^{N}\sum\limits_{m=-n}^{n}c_{nm}(k,\veg{r}'_{\rm i/s/sp})\bm{D}^{(1/2)}_{nm}(\hat{\veg{k}})\,,\label{eq:FMM2}
\end{equation}
where $\veg{k}$ is the wave vector with unit vector $\hat{\veg{k}}$ and wave number $k$. The expansion order $N$ is determined based on the size of the finest level octree boxes in the context of the multilevel fast multipole method (MLFMM)~\cite{Chew2001}. Solving the resulting linear system of equations gives the coefficients $c_{nm}$ and consequently the PWS of the TOIs. Echo suppression for interferences $\tilde{\veg{J}}_{\rm s}$ and $\tilde{\veg{J}}_{\rm sp}$ is also possible in this process by removing the corresponding PWS. The processes of the inverse source reconstructions are performed individually for different frequencies, making the approach both robust and compact. This method is highly efficient due to its implementation via the concepts of the MLFMM.

With the obtained PWS, e.g., $\tilde{\veg{J}}_{\mathrm{s}}$ for the TOIs, image generation using hierarchical disaggregation~\cite{schnattinger2012solution} is performed independently for each discrete frequency $f=1,\dots,F$
\begin{equation}
	\mathring{J}_{\mathrm{s},\,p}(k_{f},\veg{r}')=\oiint \mathcal{F}\left(\hat{\veg{k}}\cdot\hat{\veg{k}}^{\mathrm{(c)}}\right) \tilde{J}_{\mathrm{s},\,p}(\veg{k}_f,\veg{r_{\rm{s}}})\e^{-\jm \veg{k}_f \cdot (\veg{r}'-\veg{r}_{\rm{s}})} \, \dd^2 \hat{\veg{k}}\,,\label{eq:singleF}
\end{equation}
where $p \in \{x,y,z\}$ indicates component-wise operations in Cartesian coordinates. $\smash{\mathcal{F}(\hat{\veg{k}}\cdot\hat{\veg{k}}^{\mathrm{(c)}})}$ is an angular spectral windowing function that truncates undesired PWS contributions from outside the main observation area, thereby enhancing clutter suppression~\cite{eibert2024inverse}. Once the the single-frequency images $\mathring{J}_{\mathrm{s},\,p}(k_{f},\veg{r}')$ of the TOI have been obtained, they are superimposed coherently via
\begin{equation}
    J_{\mathrm{s},\,p}(\veg{r}')=\sum_{f=1}^{F}\psi_{\mathrm{s},\,p}(k_f,\veg{r}'){\mathcal{M}_{\mathrm{ s},\,p}(k_f,\veg{r}')}\,k_f\,\Delta k_f \mathring{J}_{\mathrm{s},\,p}(k_{f},\veg{r}') \,,\label{eq:sumF}
\end{equation}
where the phase and magnitude correction terms $\psi_{\mathrm{s},\,p}(k_f,\veg{r}')$ and $\mathcal{M}_{\mathrm{ s},\,p}(k_f,\veg{r}')$ are found in~\cite{wang2024TAP}. The theoretical resolution of the imaging algorithm for planar observations aligned with the coordinate system in Figure~\ref{fig:config} is estimated as~\cite{lopez-sanchez20003d}
\begin{equation}
    \delta_{x} =\frac{\lambda_{c}x_{0}}{L_{x}},\quad \delta_{z} =\frac{\lambda_{c} z_{0}}{L_{z}},\quad \text{and } \delta_{y} =\frac{\mathrm{c}_0}{B}\,,\label{eq:resolution}
\end{equation}
where $L_{x}$ and $L_{y}$ denote the aperture size, $y_{0}$ is the range distance between the aperture plane and the image domain, $B$ is the bandwidth, $\mathrm{c}_0$ is the speed of light, and $\lambda_{c}$ represents the wavelength at the center frequency.

Utilizing this imaging algorithm, the spatial position of electromagnetic scatterers, including passive tags, can be readily determined. Passive tags can cause stronger scattered fields from the TOI, providing enhanced localization capabilities, especially when imaging resources are limited, such as in terms of observation area size or signal bandwidth. They also serve as the sole beacons when the static TOIs are no longer present in the resulting images, such as when background subtraction is applied to eliminate strong clutter from the indoor environment. A requirement for passive tags in this process is that they must scatter a measurable electromagnetic field when illuminated, meaning that theoretically any sufficiently strong electromagnetic scatterer can serve as a passive tag. Another important property of passive tags is their shape, which can be used to distinguish different TOIs.

\section{Numerical Results}
\label{sec:numRes}

\subsection{Localization}

\begin{figure}[t]
	\centerline{\includegraphics[width=0.8\columnwidth,draft=false]{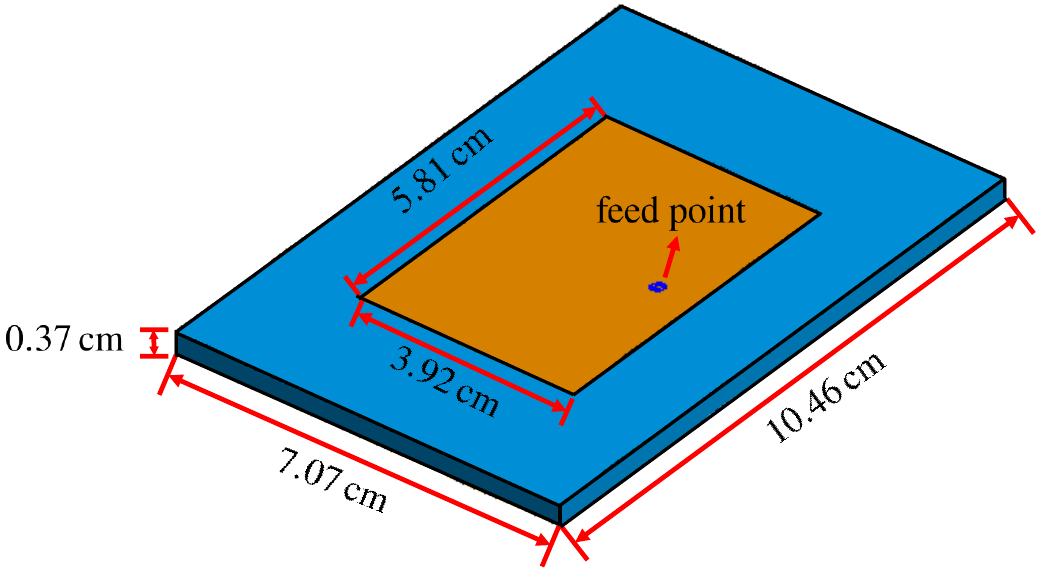}}
	\caption{Example of a patch antenna as passive tag.}
	\label{fig:patch}
\end{figure}

\begin{figure}[t]
	\centerline{\includegraphics[width=0.95\columnwidth,draft=false]{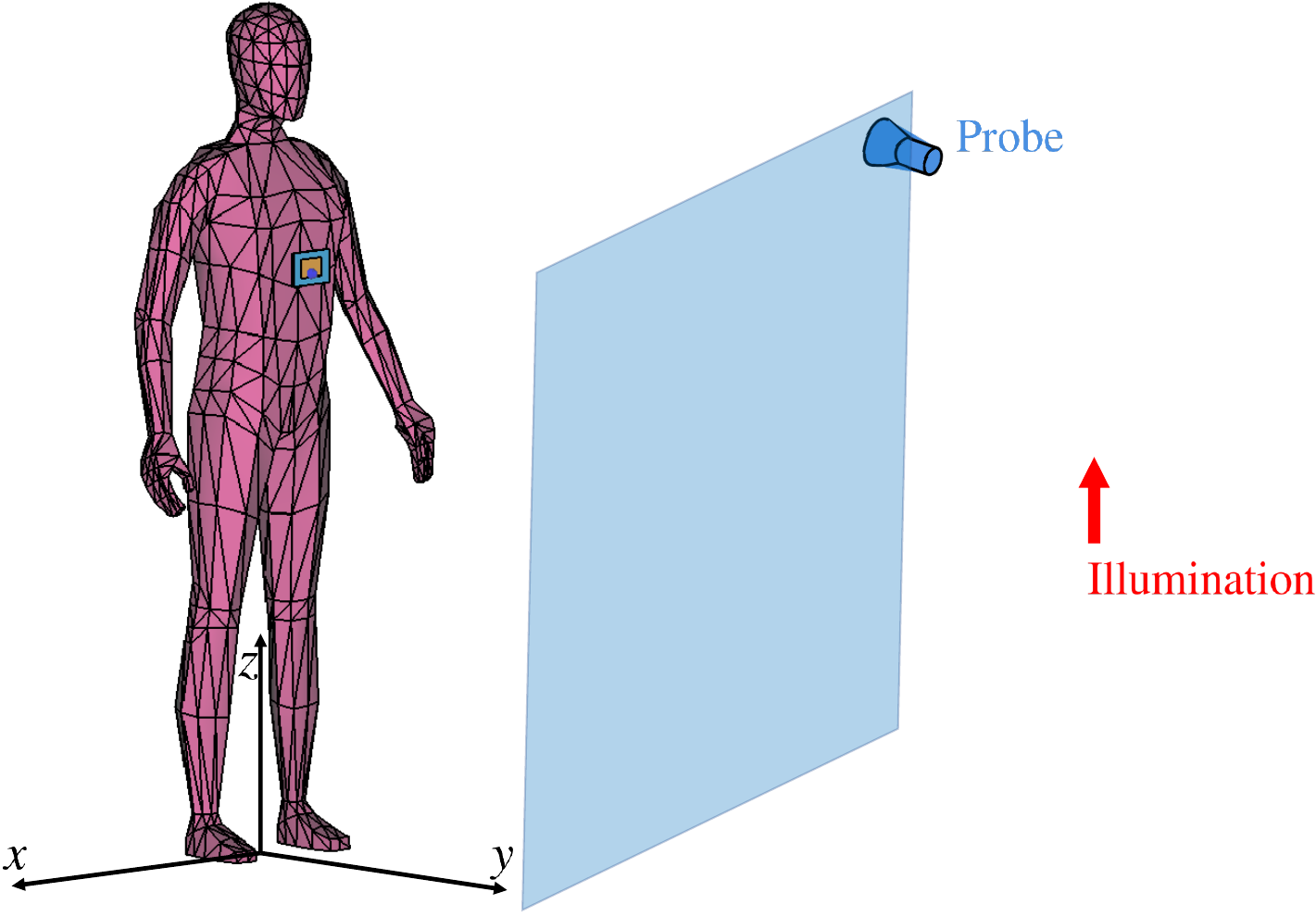}}
	\caption{Illustration of the simulation setup in FEKO including a human model and a patch antenna serving as the passive tag.}
	\label{fig:patch_man}
\end{figure}

\begin{figure}[t]
	% \centering
	% \subfloat[]{\includegraphics[scale=0.85,draft=false]{Figs/man_tag1.pdf}}%
	% \hfill
	% \subfloat[]{\includegraphics[scale=0.85,draft=false]{Figs/man_tag2.pdf}}%
	\begin{center}
        \begin{minipage}[t]{0.45\linewidth}
            \centering
            \begin{overpic}[scale=0.8,draft=false]{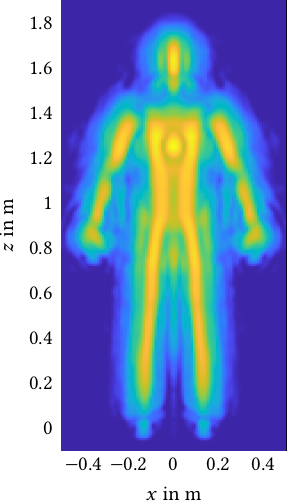}
                \put(0,102){\footnotesize{(a)}} 
            \end{overpic}
		\end{minipage}\hfill
        \begin{minipage}[t]{0.45\linewidth}
            \centering
            \begin{overpic}[scale=0.8,draft=false]{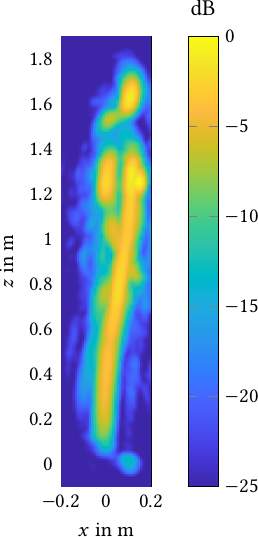}
                \put(0,95){\footnotesize{(b)}} 
            \end{overpic}
        \end{minipage}
    \end{center}
	\caption{Multi-frequency imaging result of the human body with a passive tag, where the passive tag appears as the bright spot in front of the chest of the human model.
	(a)~MIP from the front view. 
	(b)~MIP from the side view.}
	\label{fig:man_tag}
\end{figure}

\begin{figure}[t]
	% \centering
	% \subfloat[]{\includegraphics[scale=0.85,draft=false]{Figs/man_notag1.pdf}}%
	% \hfill
	% \subfloat[]{\includegraphics[scale=0.85,draft=false]{Figs/man_notag2.pdf}}%
	\begin{center}
        \begin{minipage}[t]{0.45\linewidth}
            \centering
            \begin{overpic}[scale=0.8,draft=false]{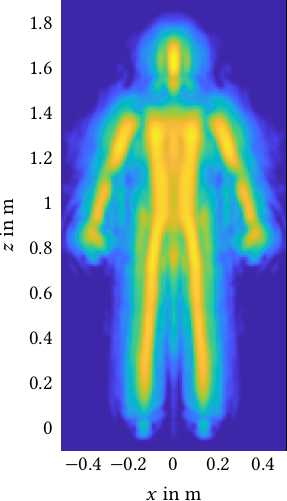}
                \put(0,102){\footnotesize{(a)}} 
            \end{overpic}
		\end{minipage}\hfill
        \begin{minipage}[t]{0.45\linewidth}
            \centering
            \begin{overpic}[scale=0.8,draft=false]{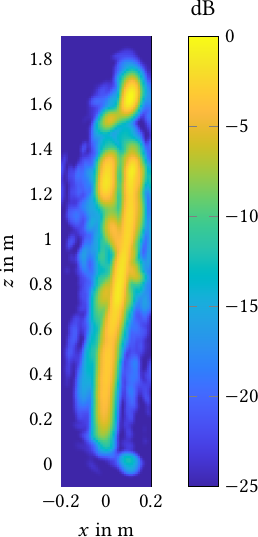}
                \put(0,95){\footnotesize{(b)}} 
            \end{overpic}
        \end{minipage}
    \end{center}
	\caption{Imaging results of the human body without the passive tag.
	(a)~MIP from the front view. 
	(b)~MIP from the side view.}
	\label{fig:man_no_tag}
\end{figure}

First, we consider a simple passive tag in the form of a patch antenna, as shown in Figure~\ref{fig:patch}. The substrate dimensions of the patch antenna are $\SI{10.46}{\centi\meter} \times \SI{7.07}{\centi\meter}$, with a thickness of $\SI{0.37}{\centi\meter}$ and a relative permittivity of 2.15. The thin patch is perfectly electrically conducting (PEC) material, with dimensions $\SI{5.81}{\centi\meter} \times \SI{3.92}{\centi\meter}$. The feed point is located at the center of the long edge of the patch, offset by $\SI{1.24}{\centi\meter}$ from the center along the short edge. The antenna is designed to resonate at a frequency of $\SI{2.4}{\giga\hertz}$, ensuring that it generates the largest possible scattered electromagnetic field within the simulation frequency range for imaging.
The patch antenna is positioned on the chest of a human model at a height of approximately $\SI{1.25}{\meter}$, as shown in Figure~\ref{fig:patch_man}. The height of the human body is around $\SI{1.8}{\meter}$. The body material is assumed to be muscle and the corresponding parameters, such as relative permittivity $\varepsilon_{r}$ and dielectric loss tangent $\tan \delta$, are frequency dependent and retrieved from an online simulator~\cite{fcc_dielectric_parameters}, e.g., $\varepsilon_{r}=52.79$ and $\tan \delta=0.24$ at $\SI{2.4}{\giga\hertz}$.

Simulations were carried out using the commercial full-wave simulation software FEKO~\cite{EMSS2023}. The human body and the passive tag were placed around the origin of the coordinate system. The illuminating source was a Hertzian dipole located at the coordinates $[x, y, z]=[0, 2, 1]\,$m, which is 2 meters in front of the body model, roughly at a height close to the abdomen. This setup aims to illuminate the whole body equally. The $x$- and $z$-components of the electric field were collected over a rectangular aperture in the plane $y=\SI{0.5}{\meter}$ extending from $[x, z]=[-3, -3]\,$m to $[x, z]=[3, 3]\,$m, which was positioned directly in front of the human body to capture all relevant scattered fields. Overall, $12\,100$ uniformly distributed probe positions on the observation plane were utilized to fulfil the desired sampling rate and ensure a proper resolution of the image. 

The Tx antenna radiates single-frequency continuous wave signals with linearly distributed frequencies from $\SI{2}{\giga\hertz}$ to $\SI{4}{\giga\hertz}$ in steps of $\SI{50}{\mega\hertz}$. After the application of the imaging algorithm and combination of the single-frequency results according to \eqref{eq:sumF}, the results in Figure~\ref{fig:man_tag} are obtained. The reconstruction area was constrained to a cubic space with the boundaries $\SI{-0.5}{\meter}\leq x \leq \SI{0.5}{\meter}$, $\SI{-0.2}{\meter}\leq y \leq \SI{-0.2}{\meter}$, and $\SI{-0.1}{\meter}\leq z \leq \SI{1.9}{\meter}$. The normalized source densities are mapped onto the faces of the cuboid enclosing the imaging region via a maximum intensity projection (MIP), where the front view and the side view are shown in Figure~\ref{fig:man_tag}(a) and (b), respectively. In addition, a simulation without the passive tag has been conducted and the resulting images are shown in Figure~\ref{fig:man_no_tag}. A comparison between Figure~\ref{fig:man_tag} and Figure~\ref{fig:man_no_tag} clearly reveals that the bright spot on the chest of the human phantom is caused by the tag.

In the imaging process, the spatial relationship between the imaged object and the observation plane is accurately reconstructed. Therefore, the location of the passive tag can be readily determined from the coordinates within the imaging domain, as illustrated Figure~\ref{fig:man_tag} and Figure~\ref{fig:man_no_tag}. However, it is important to note that the resolution of the imaging system limits the accuracy of the localization. From the side view shown in Figure~\ref{fig:man_tag}(b), the passive tag is almost merged with the human body due to the limited range resolution. The frequency range considered provides a resolution of approximately $\SI{15}{\centi\meter}$, while the closest distance between the passive tag and the surface of the human body is only $\SI{3}{\centi\meter}$.

\subsection{Identification}
Since the indoor localization method is based on the characteristics of the imaging algorithm, it also enables the identification of people or objects. This is typically challenging in conventional microwave imaging compared to optical imaging, where resolution limitations and the absence of visually recognizable color information pose difficulties in distinguishing between individuals or objects. However, in the scenario considered here, this limitation can be overcome by designing passive tags with different shapes. When the resolution is sufficient, these uniquely shaped passive tags facilitate reliable identification of people or objects.

\begin{figure*}[t]
 	\begin{center}
 	\begin{minipage}[t]{0.32\linewidth}
 		\centering
 		\begin{overpic}[height=1.1\columnwidth]{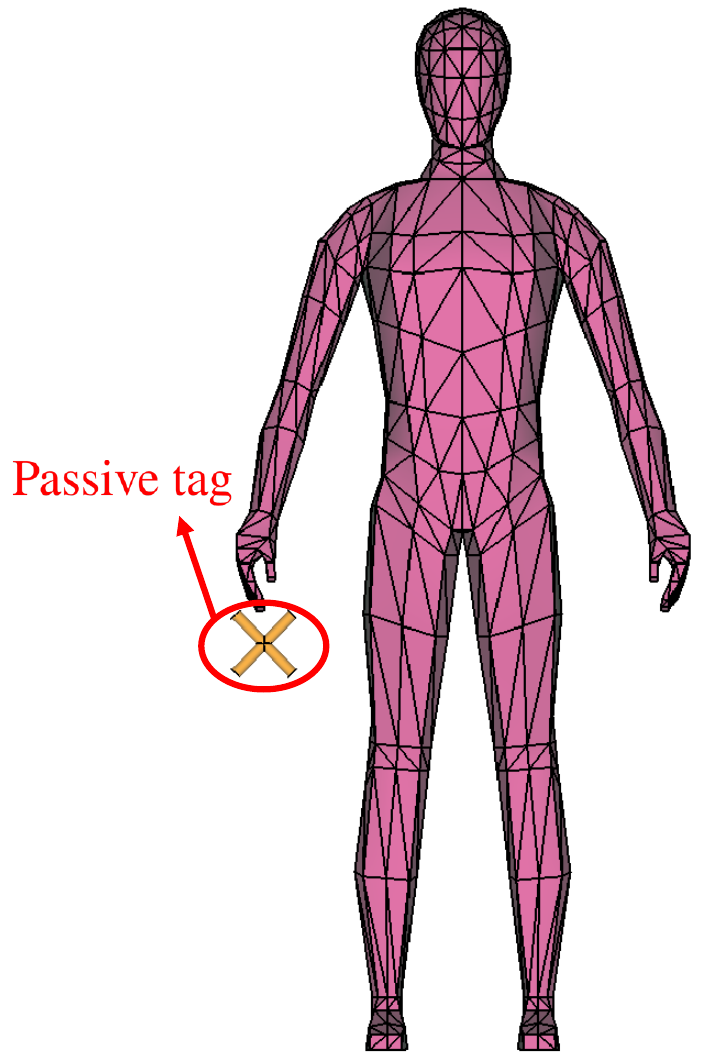}
 			\put(0,100){\footnotesize{(a)}}
 		\end{overpic}
 	\end{minipage}\hspace{-0.5cm}
 	\begin{minipage}[t]{0.32\linewidth}
 		\centering
 		\begin{overpic}[height=1.1\columnwidth]{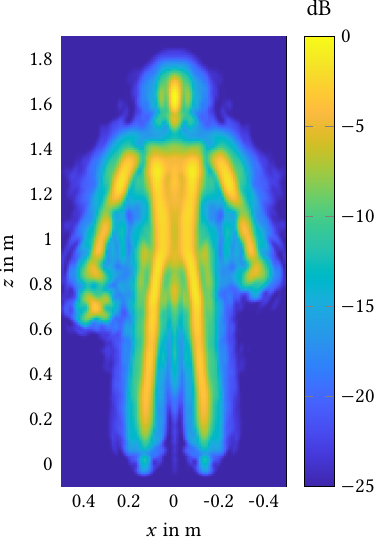}
 			\put(0,100){\footnotesize{(b)}}
 		\end{overpic}
 	\end{minipage}\hspace{0.3cm}
	 \begin{minipage}[t]{0.32\linewidth}
		\centering
		\begin{overpic}[height=1.1\columnwidth]{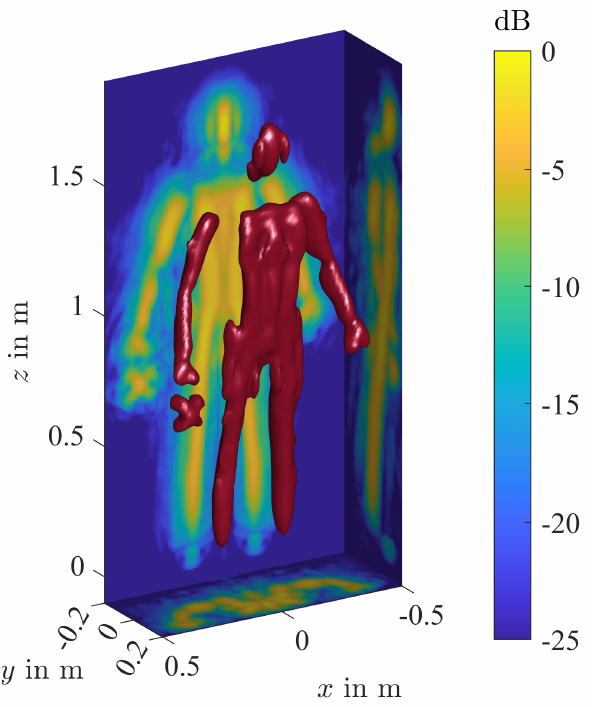}
			\put(0,100){\footnotesize{(c)}}
		\end{overpic}
	\end{minipage}\hspace{0cm}
  \end{center}
 	\caption{Illustration and imaging results of a human model and a cross-shaped PEC scatterer serving as the passive tag.
	(a)~Illustration of the simulation setup in FEKO. 
	(b)~Imaging result shown as a MIP from the front view.
	(c)~Imaging result presented as a combination of an isosurface at a threshold value of $\SI{-10}{\deci\bel}$ and 2-D MIPs from different views.}
 	\label{fig:tag_cross}
\end{figure*}

To demonstrate this, several simple PEC scatterers with different shapes were used as passive tags replacing the previously discussed patch antenna. These new passive tags were simulated alongside the human model using FEKO. As shown in Figure~\ref{fig:tag_cross}(a), a cross-shaped structure was positioned directly below the right hand of the human model. With all other simulation settings remaining unchanged, the MIP from the front view is presented in Figure~\ref{fig:tag_cross}(b). In addition, a 3-D display using an isosurface at a threshold value of $\SI{-10}{\deci\bel}$ combined with 2-D MIPs from different views is provided in Figure~\ref{fig:tag_cross}(c). Both the human model and the cross-shaped structure beneath the hand are clearly reconstructed in these imaging results. The geometrical characteristics of the tags can serve as unique identifiers for individuals. By having different people wear distinct types of tags, it becomes possible to identify different individuals. 

\begin{figure*}[t]
	\begin{center}
	\begin{minipage}[t]{0.48\linewidth}
		\centering
		\begin{overpic}[width=0.7\columnwidth]{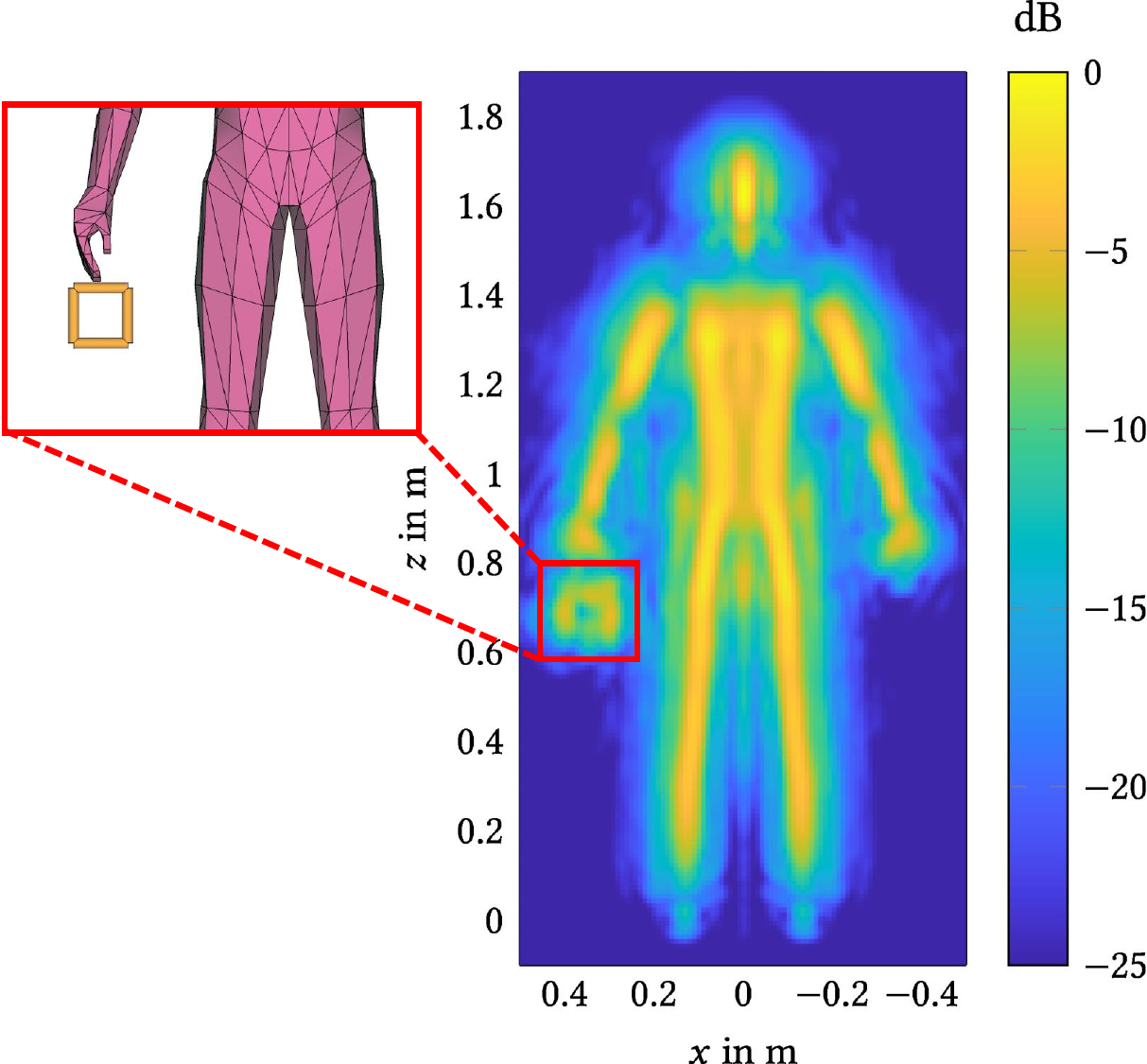}
			\put(-5,95){\footnotesize{(a)}}
		\end{overpic}
	\end{minipage}\hspace{0cm}
	\begin{minipage}[t]{0.48\linewidth}
		\centering
		\begin{overpic}[width=0.7\columnwidth]{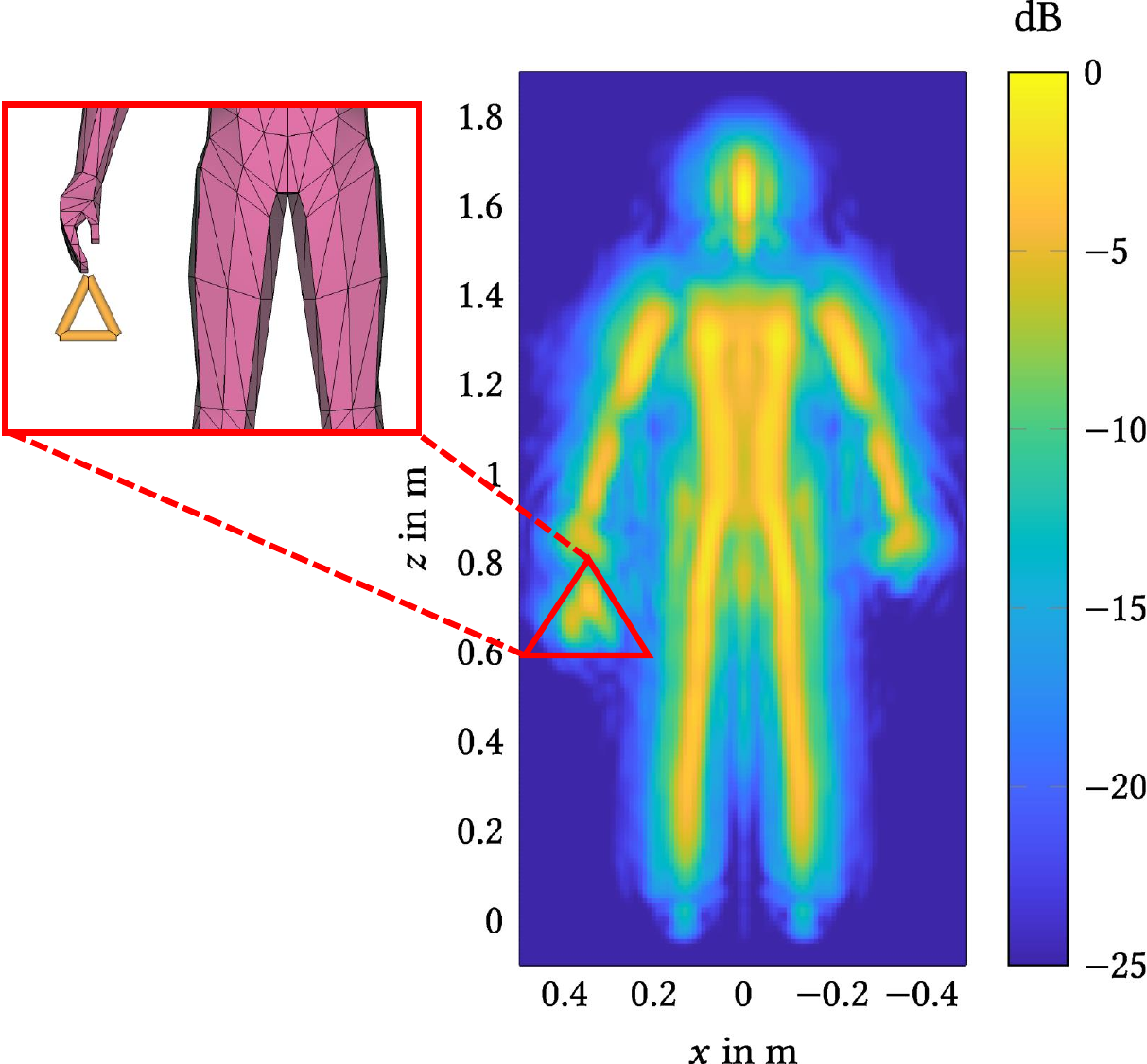}
			\put(-5,95){\footnotesize{(b)}}
		\end{overpic}
	\end{minipage}
 \end{center}
	\caption{Imaging results of the human body with different shapes of passive tags, including zoomed-in views for the tags.
		(a)~Square-shaped passive tag. 
		(b)~Triangular-shaped passive tag.}
	\label{fig:tag_tri_rec}
\end{figure*}

\begin{figure}[t]
   \centerline{\includegraphics[width=0.95\columnwidth,draft=false]{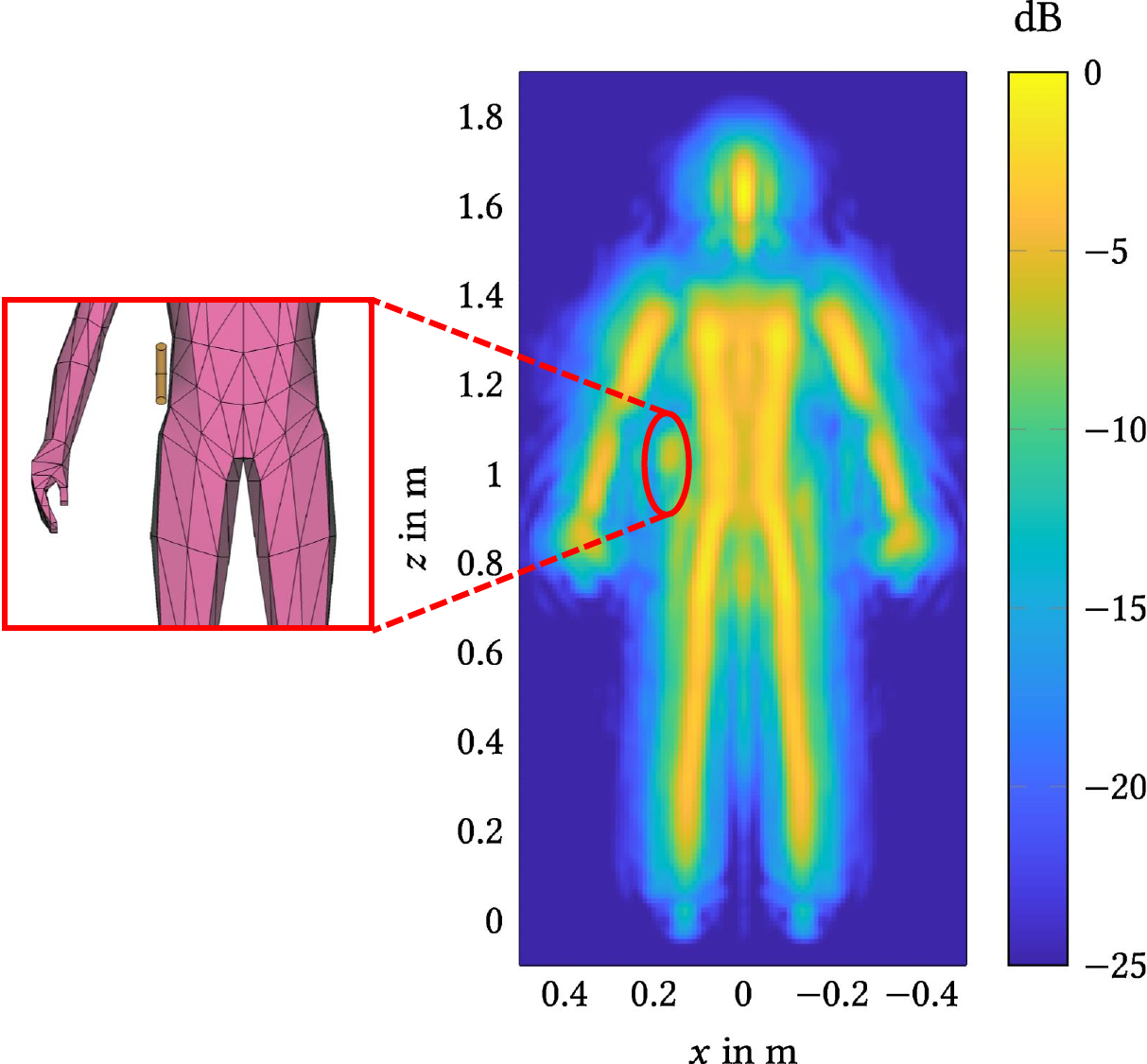}}
   \caption{Imaging result of the cross-shaped passive tag positioned near the waist of the human body, together with a zoomed-in view of the tag. The tag is less visible due to weaker scattered fields.}
   \label{fig:tag_cross_hide}
\end{figure}
Further simulations and imaging results are presented in Figure~\ref{fig:tag_tri_rec}, where Figure~\ref{fig:tag_tri_rec}(a) shows a square-shaped passive tag, and Figure~\ref{fig:tag_tri_rec}(b) shows a triangular-shaped passive tag. All these imaging results successfully reveal the shape of the tags. The size of the tags was designed to be large enough to be recognized under the resolution constraints imposed by the imaging configurations, as given by \eqref{eq:resolution}. As resolution increases, it becomes possible to design smaller and more portable tags. 

Regarding privacy concerns, it is essential to ensure that the localization and identification system fulfills ethical requirements and that sensitive data is not misused. However, this can be challenging, especially in cases where there is a lack of trust between users and service providers. The considered method, based on electromagnetic scattering and imaging, offers a privacy protection mechanism by providing users with full control over customizable settings, enabling a tailored localization algorithm. A straightforward objective in this regard is reducing the intensity of the electromagnetic field scattered by the passive tag. For instance, users can simply rotate the tag or place it differently as illustrated in Figure~\ref{fig:tag_cross_hide}. The cross-shaped passive tag is here placed close to the waist of the human bod. The corresponding imaging results clearly show that the scattering strength of the passive tag is significantly reduced and the identification becomes impossible in this scenario. A similar concept can be applied to the manufacturing of passive tags. By designing the tags to alter their electromagnetic scattering properties, users are granted full control and flexibility in determining how the tags behave. This allows for a customizable approach where the scattering intensity and visibility can be adjusted according to user preferences, thereby enhancing both privacy and control over the localization system.

A potential application of the discussed localization and identification method, similar to the case mentioned in~\cite{keskin2018localization}, involves installation in a museum room containing various exhibits. The system can detect visitors and, based on their estimated location, determine which exhibit they are interested in. Furthermore, with identification capabilities, customized services such as explanations in different languages are possible. The utilized resource such as aperture size and bandwidth is adjustable according to the required localization accuracy. In such a scenario, accuracy requirements are less stringent and fewer resources are sufficient to fulfill the needs of the systems.

% \begin{figure}[t]
% 	\centering
% 	% \subfloat[]{\input{Figures/absorber_cut1.tex}}%
% 	\subfloat[]{\includegraphics[scale=0.25,draft=false]{Figs/tag_rec.pdf}}%
% 	% \subfloat[]{\includegraphics[width=1.7in,height=1.0625in]{Figures/absorber_cut1.pdf}}%
% 	\hfill
% 	% \subfloat[]{\input{Figures/absorber_cut2.tex}}%
% 	\subfloat[]{\includegraphics[scale=0.25,draft=false]{Figs/tag_tri.pdf}}%
% 	% \subfloat[]{\includegraphics[width=1.7in,height=1.0625in]{Figures/absorber_cut2.pdf}}%
% 	\caption{Imaging results of the human body with different shapes of passive tags. 
% 	(a)~Square-shaped passive tag. 
% 	(b)~Triangular-shaped passive tag.}
% 	\label{fig:tag_tri_rec}
% \end{figure}

\section{Measurements Results}
\label{sec:meaRes}
\begin{figure}[t]
	\centerline{\includegraphics[width=0.95\columnwidth,draft=false]{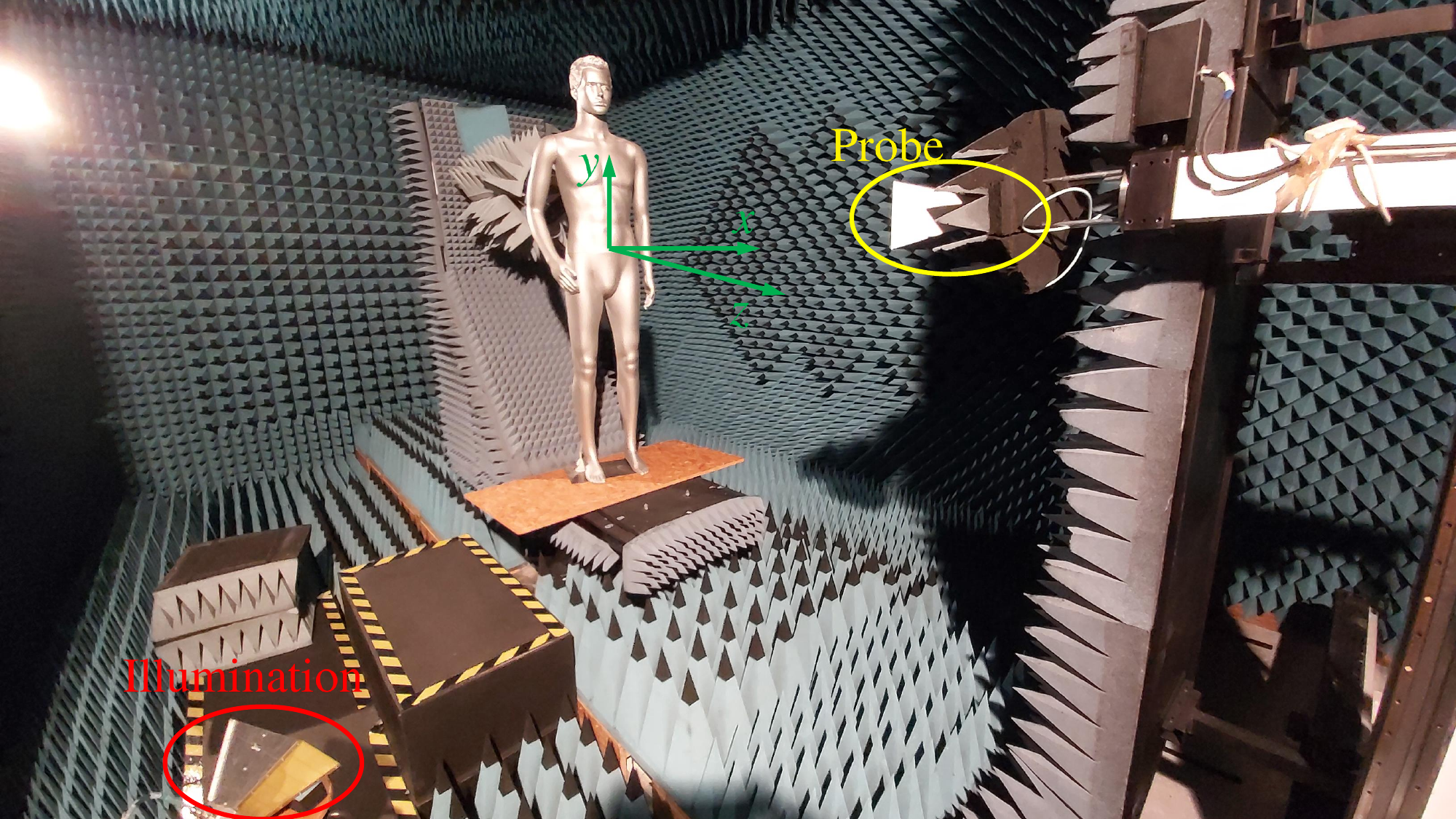}}
	\caption{Measurement configuration utilized in the anechoic chamber. The origin of the coordinate system is set close to the waist of the mannequin.}
	\label{fig:measurement}
\end{figure}
\begin{figure}[t]
	% \centering
	% \subfloat[]{\includegraphics[width=0.45\columnwidth,draft=false]{Figs/mannequin_patch.pdf}}%
	% \hfill
	% \subfloat[]{\includegraphics[width=0.45\columnwidth,draft=false]{Figs/photo_patch}}%
	\begin{center}
        \begin{minipage}[t]{0.45\linewidth}
            \centering
            \begin{overpic}[width=\linewidth,draft=false]{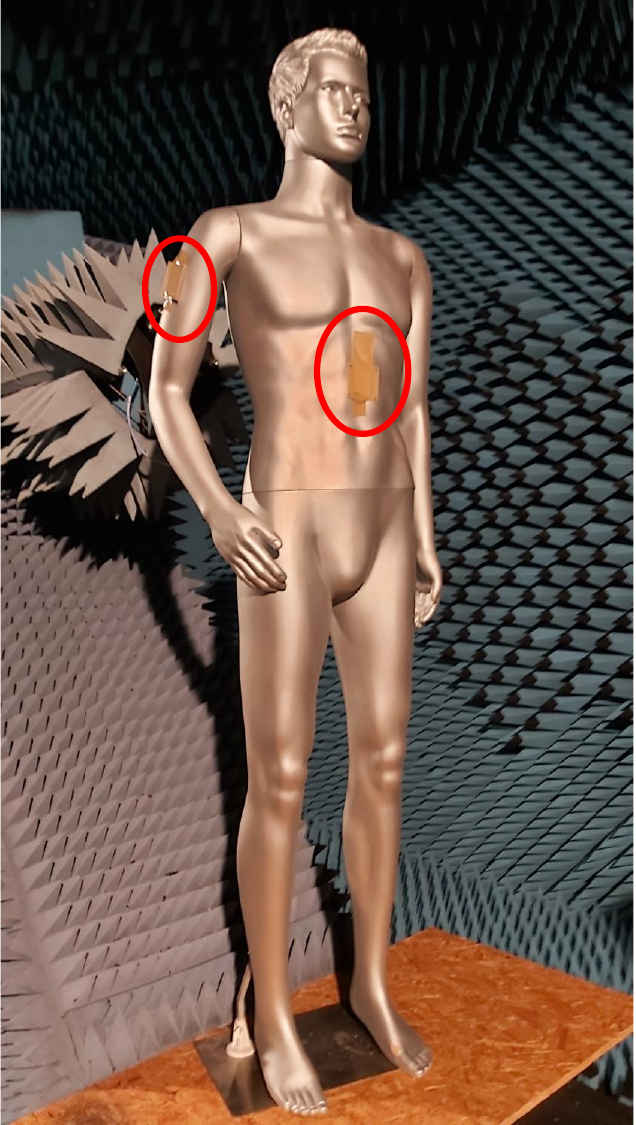}
                \put(-5,102){\footnotesize{(a)}} 
            \end{overpic}
		\end{minipage}\hfill
        \begin{minipage}[t]{0.45\linewidth}
            \centering
            \begin{overpic}[width=\linewidth,draft=false]{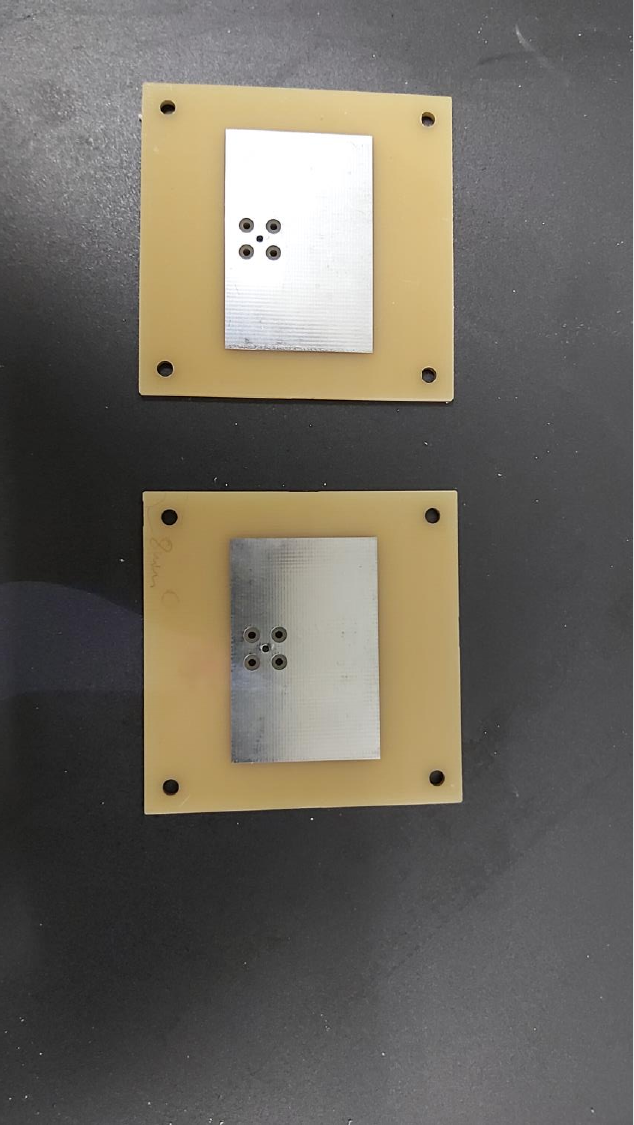}
                \put(-5,102){\footnotesize{(b)}} 
            \end{overpic}
        \end{minipage}
    \end{center}
	\caption{(a)~Photograph of the mannequin with two passive tags.
			(b)~Close-up image of the two patch antennas serving as the passive tags in the measurement.}
	\label{fig:mea_man_patch}
\end{figure}
\begin{figure*}[t]
	\begin{center}
	\begin{minipage}[t]{0.32\linewidth}
		\centering
		\begin{overpic}[height=1\columnwidth]{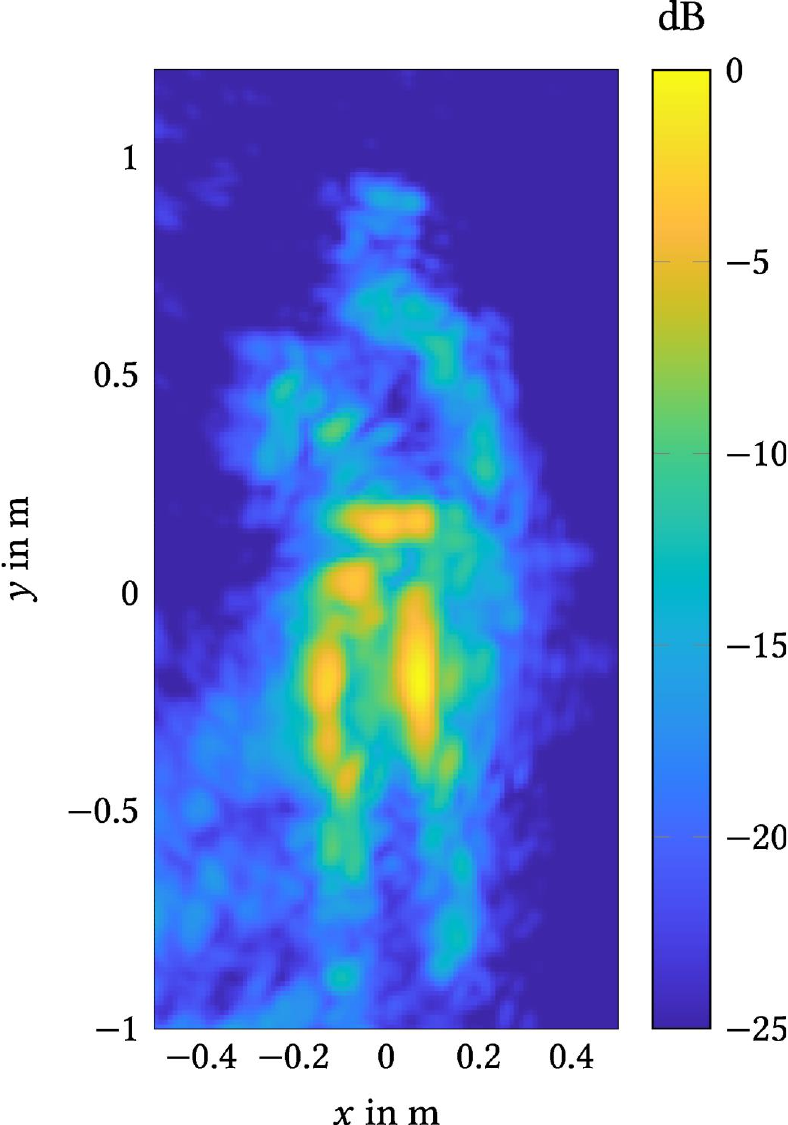}
			\put(0,100){\footnotesize{(a)}}
		\end{overpic}
	\end{minipage}\hspace{0cm}
	\begin{minipage}[t]{0.32\linewidth}
		\centering
		\begin{overpic}[height=1\columnwidth]{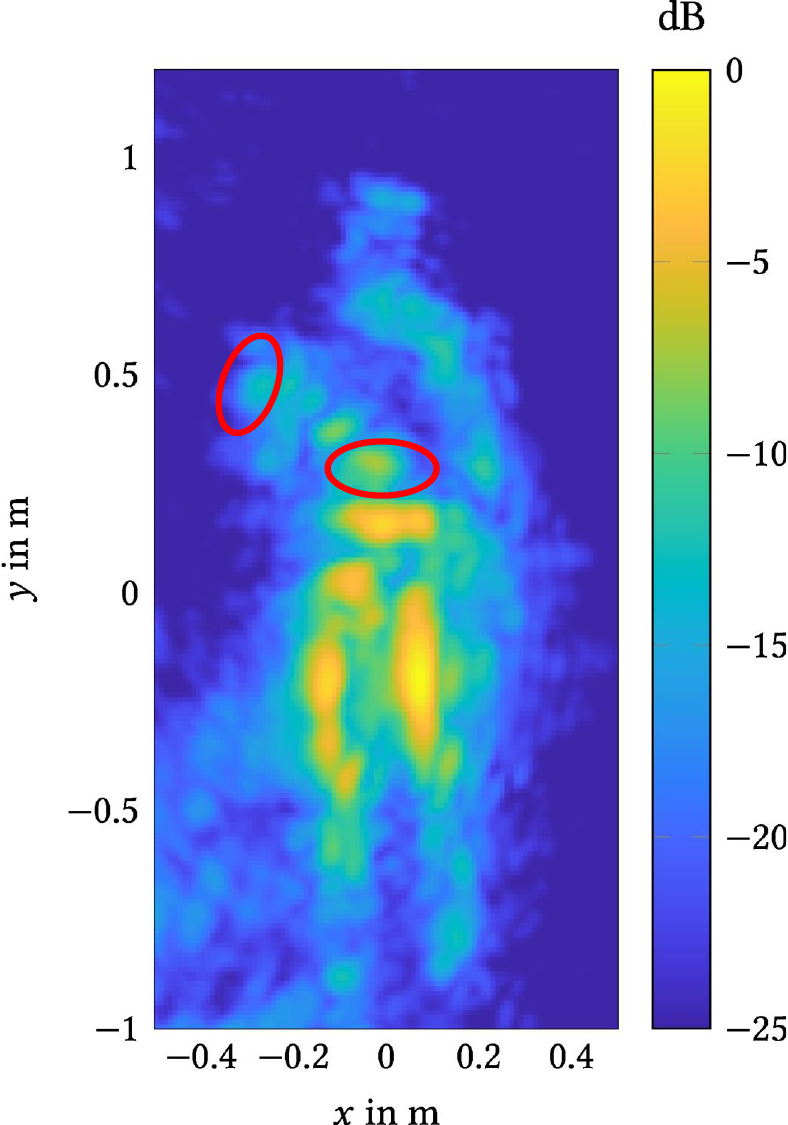}
			\put(0,100){\footnotesize{(b)}}
		\end{overpic}
	\end{minipage}\hspace{0cm}
	\begin{minipage}[t]{0.32\linewidth}
	   \centering
	   \begin{overpic}[height=1\columnwidth]{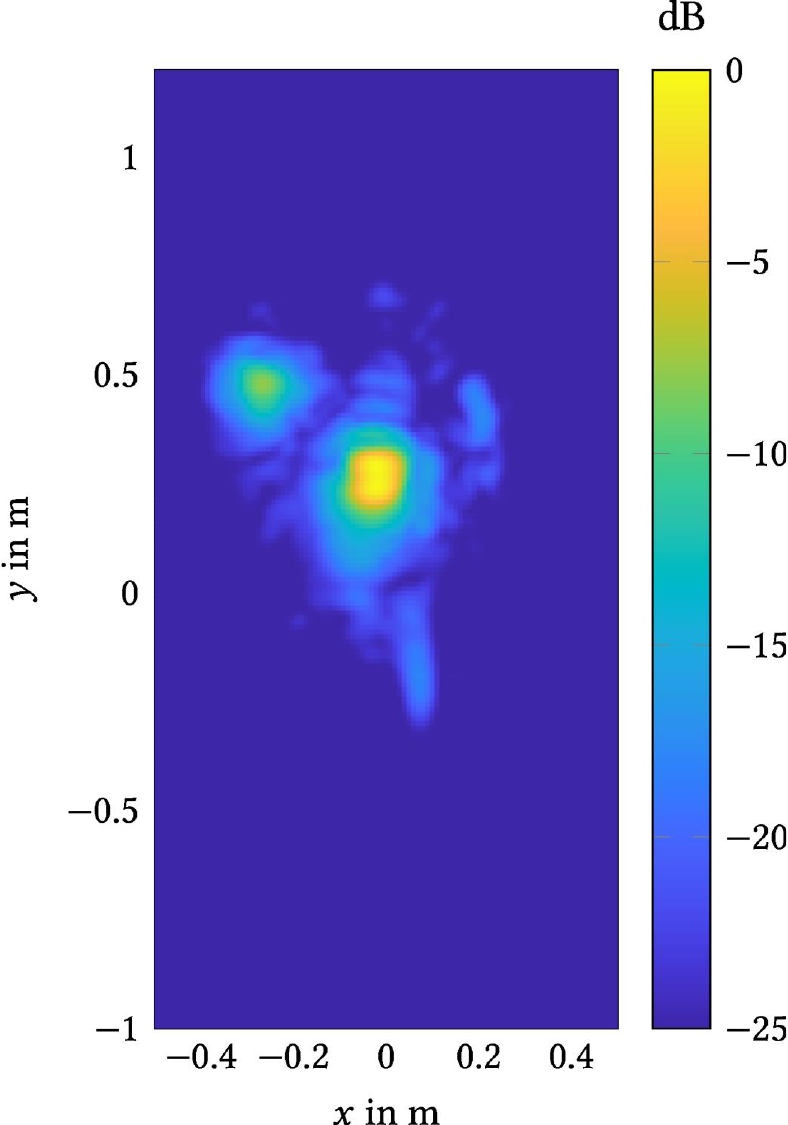}
		   \put(0,100){\footnotesize{(c)}}
	   \end{overpic}
   \end{minipage}\hspace{0cm}
 \end{center}
	\caption{Imaging results of the mannequin with or without passive tags shown as a MIP from the front view.
   (a)~Without passive tags. 
   (b)~With two passive tags and the positions of the tags highlighted by red circles.
   (c)~Imaging result after background subtraction.}
	\label{fig:img_man_patch}
\end{figure*}
\begin{figure*}[t]
	\begin{center}
	\begin{minipage}[t]{0.32\linewidth}
		\centering
		\begin{overpic}[height=1\columnwidth]{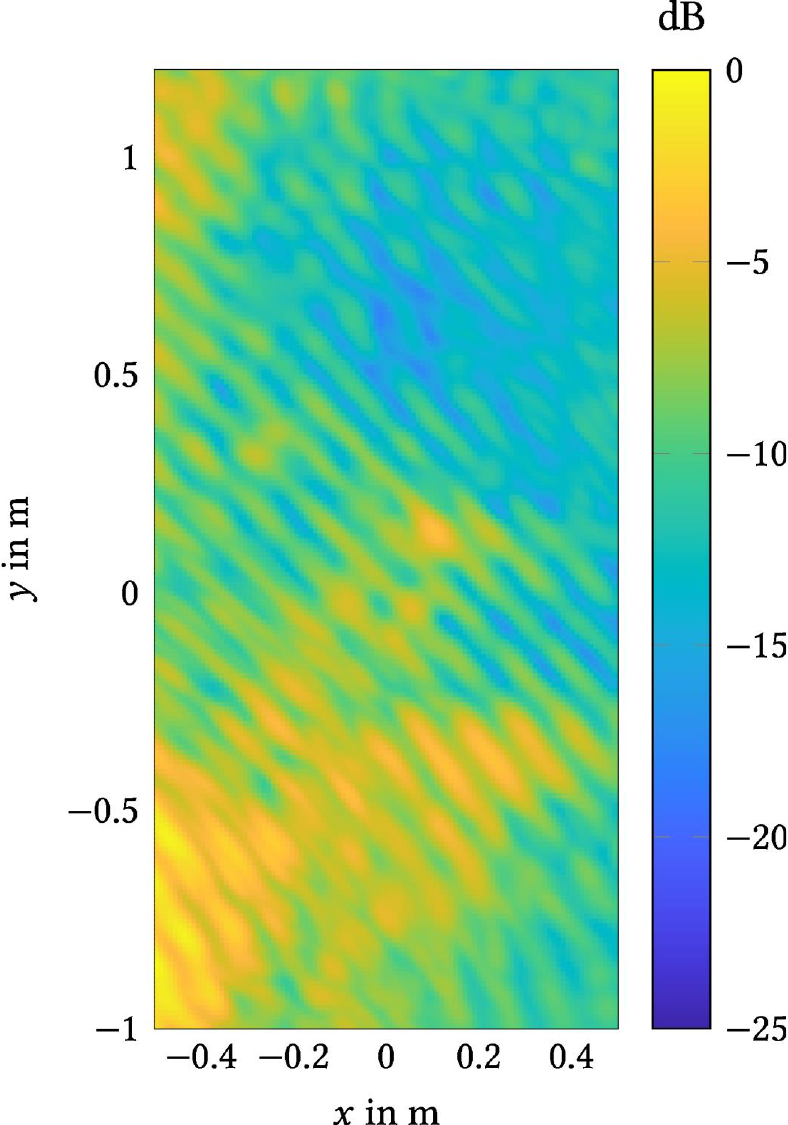}
			\put(0,100){\footnotesize{(a)}}
		\end{overpic}
	\end{minipage}\hspace{0cm}
	\begin{minipage}[t]{0.32\linewidth}
		\centering
		\begin{overpic}[height=1\columnwidth]{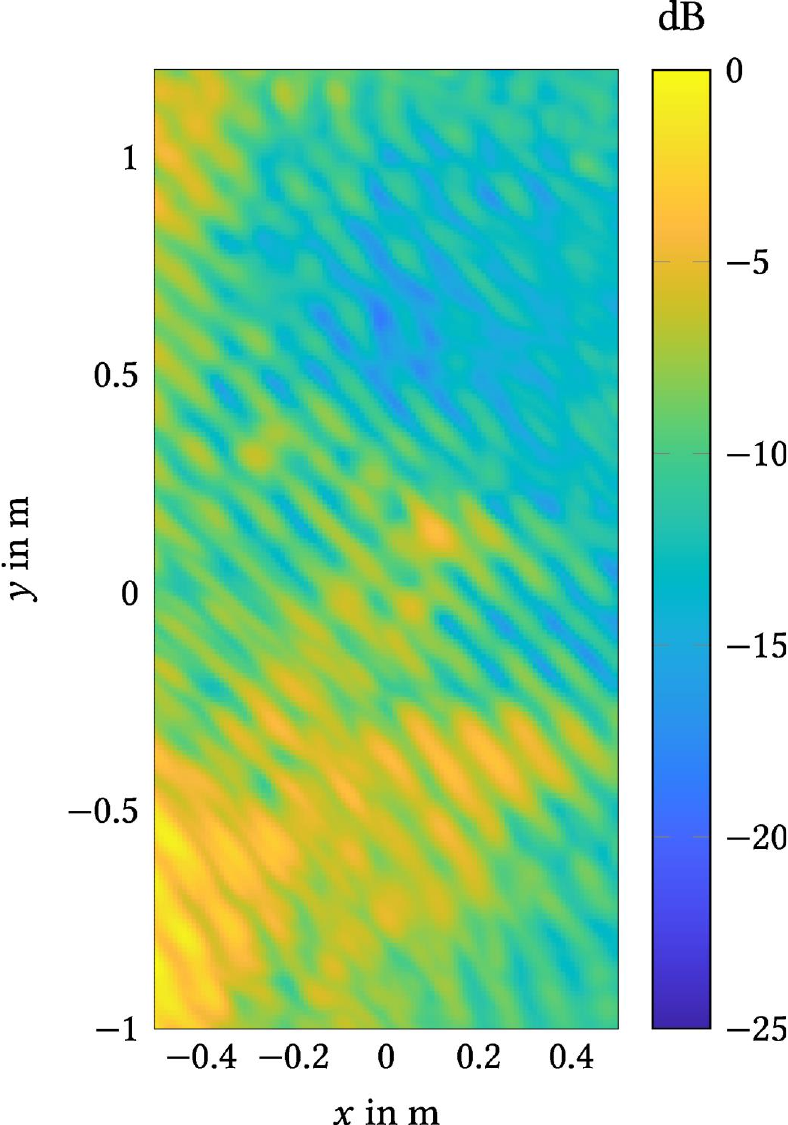}
			\put(0,100){\footnotesize{(b)}}
		\end{overpic}
	\end{minipage}\hspace{0cm}
	\begin{minipage}[t]{0.32\linewidth}
	   \centering
	   \begin{overpic}[height=1\columnwidth]{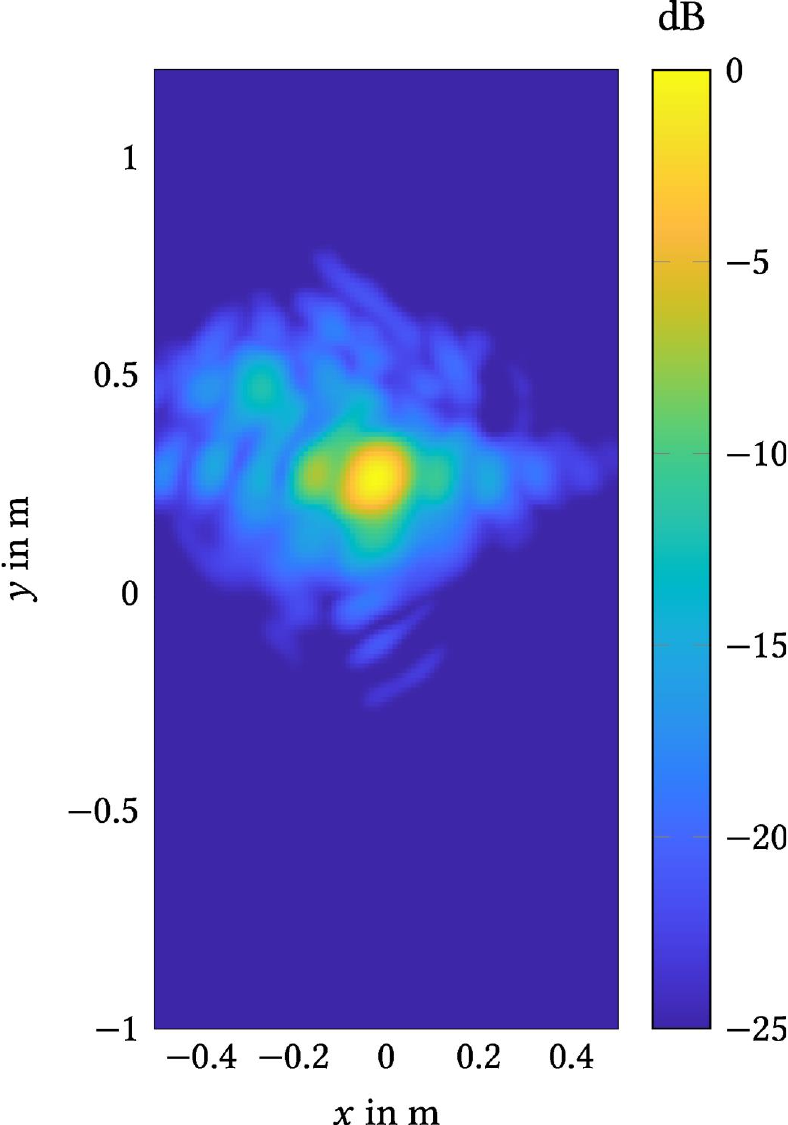}
		   \put(0,100){\footnotesize{(c)}}
	   \end{overpic}
   \end{minipage}\hspace{0cm}
 \end{center}
	\caption{Imaging results of the mannequin with or without passive tags shown as a MIP from the front view. The frequency ranges from $\SI{2.2}{\giga\hertz}$ to $\SI{2.4}{\giga\hertz}$ with five equidistant frequency points.
	(a)~Without passive tags. 
	(b)~With two passive tags.
	(c)~After background subtraction.}
	\label{fig:img_narrowband}
\end{figure*}
\begin{figure*}[t]
	\begin{center}
	\begin{minipage}[t]{0.32\linewidth}
		\centering
		\begin{overpic}[height=1\columnwidth]{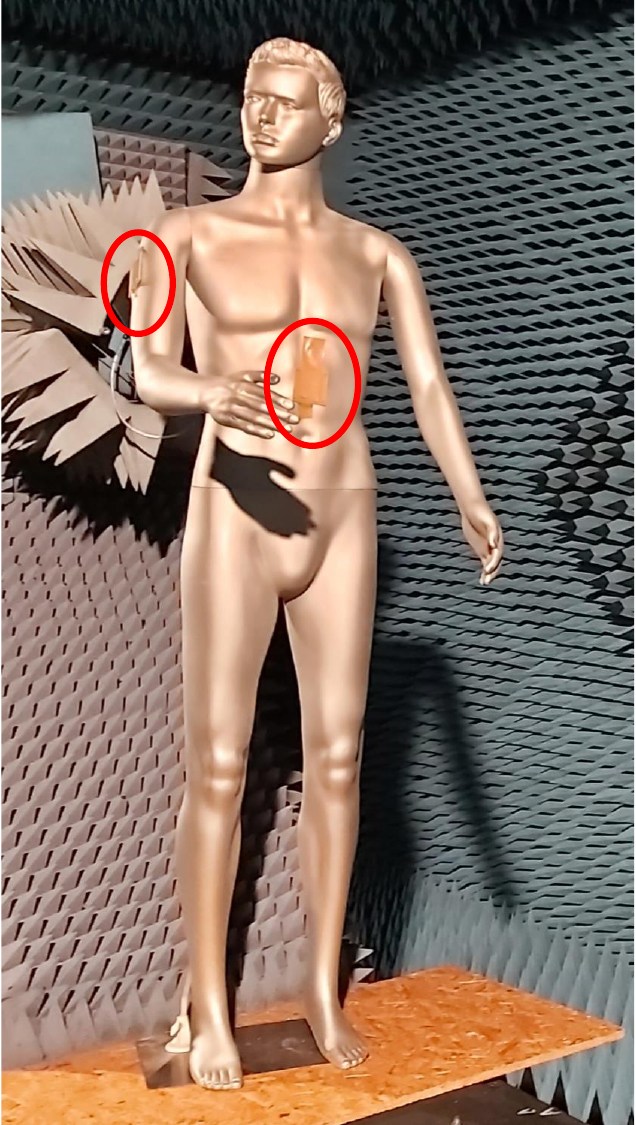}
			\put(-5,102){\footnotesize{(a)}}
		\end{overpic}
	\end{minipage}\hspace{0cm}
	\begin{minipage}[t]{0.32\linewidth}
		\centering
		\begin{overpic}[height=1\columnwidth]{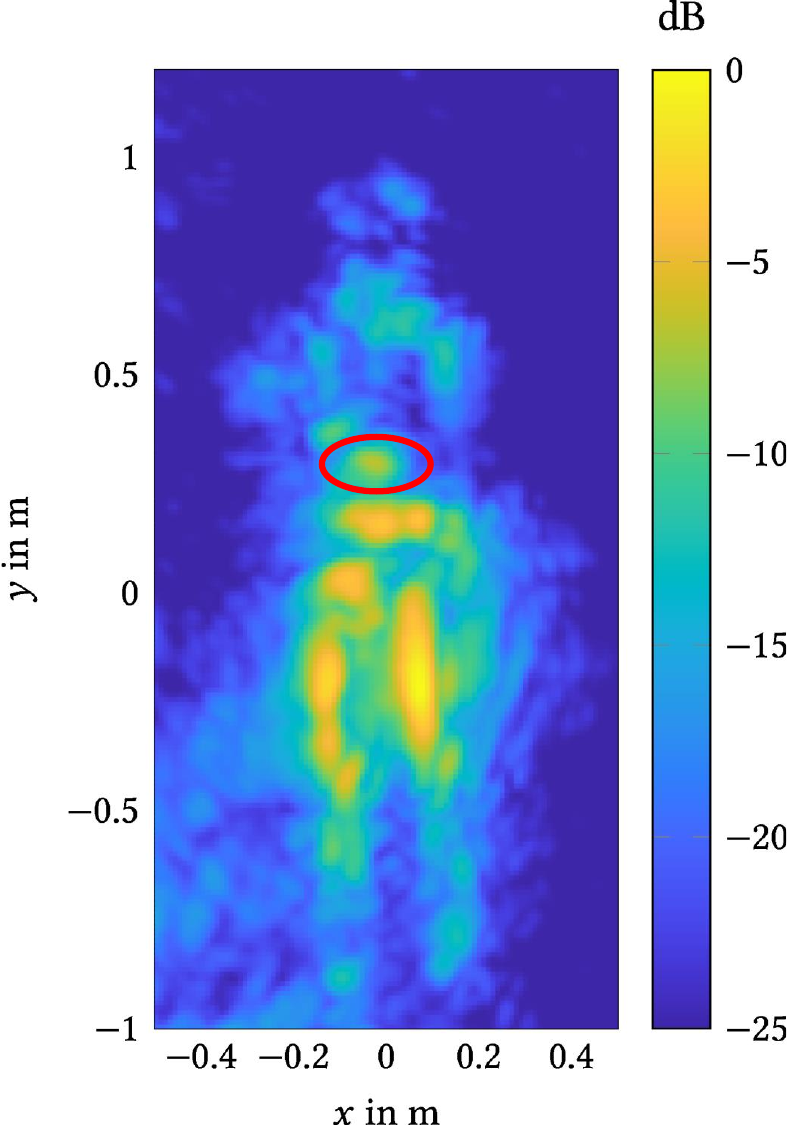}
			\put(-5,102){\footnotesize{(b)}}
		\end{overpic}
	\end{minipage}\hspace{0cm}
	\begin{minipage}[t]{0.32\linewidth}
	   \centering
	   \begin{overpic}[height=1\columnwidth]{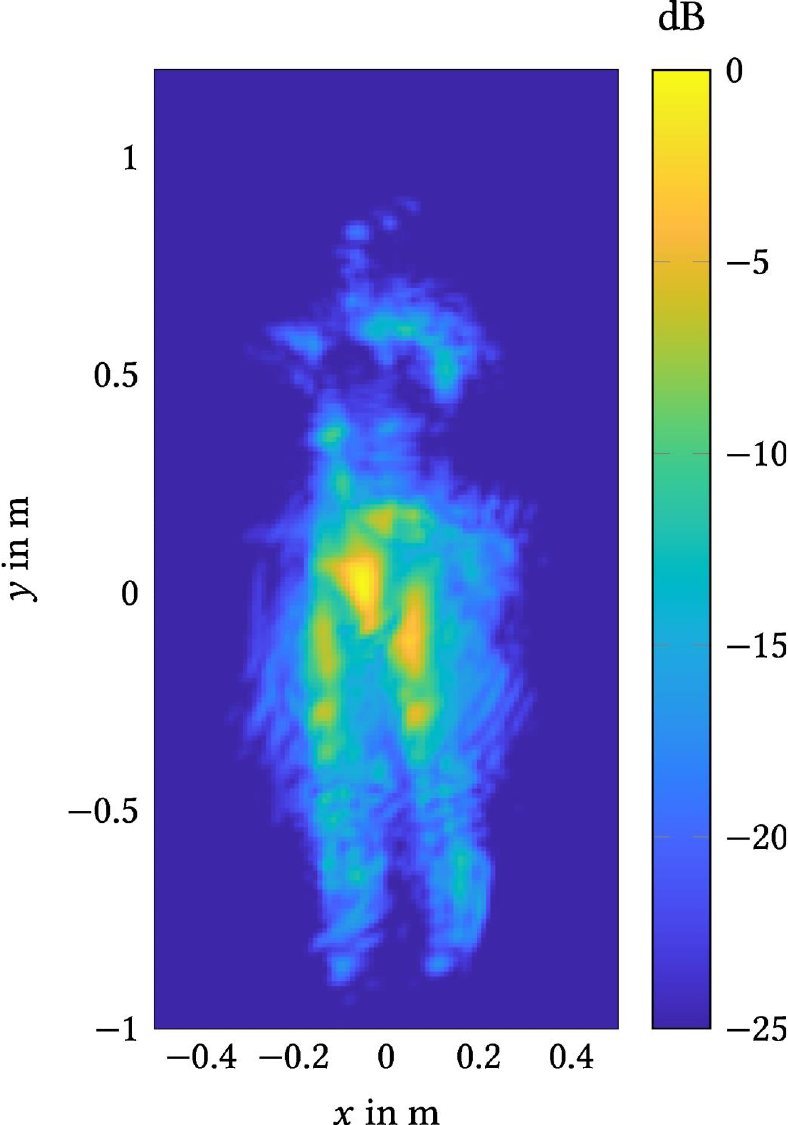}
		   \put(-5,102){\footnotesize{(c)}}
	   \end{overpic}
   \end{minipage}\hspace{0cm}
 \end{center}
	\caption{
   (a)~Photograph of the mannequin with two passive tags, with the head and arms adjusted from the previous measurement.
   (b)~Imaging result over a frequency bandwidth of $\SI{1.5}{\giga\hertz}$ to $\SI{6}{\giga\hertz}$.
   (c)~Imaging result over a frequency bandwidth of $\SI{6}{\giga\hertz}$ to $\SI{12}{\giga\hertz}$.}
	\label{fig:img2_man_patch}
\end{figure*}

% \begin{figure*}[t]
% 	\begin{center}
% 	\begin{minipage}[t]{0.24\linewidth}
% 		\centering
% 		\begin{overpic}[height=1.35\columnwidth]{Figs/mannequin_patch.pdf}
% 			\put(-6,95){\footnotesize{(a)}}
% 		\end{overpic}
% 	\end{minipage}\hspace{0cm}
% 	\begin{minipage}[t]{0.24\linewidth}
% 		\centering
% 		\begin{overpic}[height=1.35\columnwidth]{Figs/photo_patch}
% 			\put(-6,95){\footnotesize{(b)}}
% 		\end{overpic}
% 	\end{minipage}
% 	\begin{minipage}[t]{0.24\linewidth}
% 		\centering
% 		\begin{overpic}[height=1.35\columnwidth]{Figs/mea_man_notag.pdf}
% 			\put(-6,95){\footnotesize{(c)}}
% 		\end{overpic}
% 	\end{minipage}\hspace{0cm}
% 	\begin{minipage}[t]{0.24\linewidth}
% 		\centering
% 		\begin{overpic}[height=1.35\columnwidth]{Figs/mea_man_tag.pdf}
% 			\put(-6,95){\footnotesize{(d)}}
% 		\end{overpic}
% 	\end{minipage}
%  \end{center}
% 	\caption{ 
% 		(a)~Photograph of the mannequin with two passive tags.
% 		(b)~Close-up image of two patch antennas serving as the passive tags in the measurement.
% 		(c)~Imaging result of the mannequin without passive tags shown as a MIP from the front view.
% 		(d)~Imaging result of the mannequin with two passive tags shown as a MIP from the front view, with the positions of the tags highlighted by red circles.}
% 	\label{fig:mea_img_patch}
% \end{figure*}

Several measurement campaigns were conducted in the anechoic antenna measurement chamber at the Technical Technical University of Munich. A mannequin painted with zinc-aluminum spray was placed and fixed on the spherical positioner, while two linearly polarized double-ridged horn antennas (of type HF906 from Rohde \& Schwarz \cite{HF906} and of type DRH18 from RFspin \cite{Drh18}) were utilized as the illuminating antenna and the scanning probe, respectively, as depicted in Figure~\ref{fig:measurement}. The probe was mounted on a planar scanner, which is capable of mechanically moving along the $x$- and $y$-direction in the plane $z=\SI{1.8}{\meter}$ extending from $[x, y]=[-1.2, -1]\,$m to $[x, y]=[1.2, 1.2]\,$m. Observation data for both $x$- and $y$-polarization were collected at $11\,110$ distinct positions uniformly distributed across the measurement plane. 

The first measurement was performed without any passive tag as a reference. In the second measurement, two patch antennas were attached to the right arm and abdomen of the mannequin using tape, as shown in Figure~\ref{fig:mea_man_patch}(a) and highlighted by red circles. The patch antennas are narrowband with a resonance frequency of approximately $\SI{2.425}{\giga\hertz}$. A close-up image of the antennas is provided in Figure~\ref{fig:mea_man_patch}(b). Both measurements were conducted via a vector network analyzer in a stepped-frequency mode for frequencies ranging from $\SI{1.5}{\giga\hertz}$ to $\SI{6}{\giga\hertz}$ with a step size of $\SI{50}{\mega\hertz}$. While such measurements deliver directly the single-frequency observation data for the subsequent imaging process, in a realistic deployment of the method with ubiquitous Tx signals of a certain bandwidth, the received signals are typically measured with a wideband receiver and Fourier transformed in order to provide the required single-frequency observation data.

The results obtained with and without passive tags are presented in Figure~\ref{fig:img_man_patch}(a) and (b), respectively. The visualization domains are within a cubic space bounded by $\SI{-0.5}{\meter}\leq x \leq \SI{0.5}{\meter}$, $\SI{-1}{\meter}\leq y \leq \SI{1.2}{\meter}$, and $\SI{-0.1}{\meter}\leq z \leq \SI{0.4}{\meter}$. Due to the non-uniform illumination of the mannequin by the transmitter, the image of the mannequin shows stronger induced equivalent sources in the region directly facing the illumination source, particularly around the waist. Nonetheless, the overall shape of the human body has been successfully reconstructed. In addition, two bright spots are evident at the locations where the antennas are placed, serving as beacons to determine the precise coordinates of the passive tags. These bright spots become even clearer when the result in Figure~\ref{fig:img_man_patch}(a) is used as a background and subtracted from Figure~\ref{fig:img_man_patch}(b), making the positions of the two antennas more pronounced, as shown in Figure~\ref{fig:img_man_patch}(c). 

Due to the narrowband radiation characteristics of the antennas, the strength of the back-scattered fields is insufficient across the entire measured frequency band. As a result, the imaging performance for the mannequin in Figure~\ref{fig:img_man_patch} is superior to that of the passive tags. 
% Specifically, the antennas resonate at approximately $\SI{2.425}{\giga\hertz}$ while the measurements are performed from $\SI{1.5}{\giga\hertz}$ to $\SI{6}{\giga\hertz}$.
On the other hand, this specific frequency behavior is also advantageous for privacy protection, particularly when imaging of the TOI itself such as the mannequin in this case, is not desirable. This can be achieved by limiting the imaging resources. Restricting the frequency bandwidth reduces resolution, thereby decreasing imaging performance. Note that more image artifacts are usually encountered when a smaller number of single-frequency images is superimposed~\cite{wang2024TAP}. 
% Indeed, successful imaging reconstruction of the mannequin itself depends on the superposition of a certain number of single-frequency images. 
When the imaging resources are constrained, imaging for complex or weak scattering objects, e.g., the mannequin, becomes challenging. However, it remains possible to localize passive tags. For instance, when only five frequencies spanning from $\SI{2.2}{\giga\hertz}$ to $\SI{2.4}{\giga\hertz}$ are employed for the setup of Figure~\ref{fig:img_man_patch}, the imaging results in Figure~\ref{fig:img_narrowband} are obtained. As observed, the entire visualization domain is dominated by artifacts, lacking any discernible shape features, and appears to be nearly identical regardless of the presence of the tags. Solely looking at Figure~\ref{fig:img_narrowband}(a) and (b), a clear localization of the tags is not possible. However, after background subtraction, the tag on the abdomen remains clearly visible and can still be utilized for localization, while the visibility of the tag on the arm of the mannequin is worse, as shown in Figure~\ref{fig:img_narrowband}(c). Background subtraction can also effectively remove clutter due to static objects such as walls and furniture in indoor environments~\cite{dubey2024reconciling}. In this case, it also enhances the robustness of the localization method under limited imaging resources. Additionally, it enables the removal of undesired static TOIs from the imaging results, thereby supporting privacy protection.

The frequency selective property enhances the flexibility of the localization system and aids in building a privacy protection mechanism. For instance, two distinct localization systems operating at different frequencies can be installed in the same room and utilized for different TOIs and purposes, while maintaining privacy isolation from each other. In order to illustrate the effect of a narrowband tag and its frequency selective behavior, two additional measurements were conducted using different frequency bands and the results are shown in Figure~\ref{fig:img2_man_patch}. The head and arms of the mannequin were slightly adjusted while the passive tags remained unchanged. The imaging results for the frequency ranges of $\SI{1.5}{\giga\hertz}$ to $\SI{6}{\giga\hertz}$ and $\SI{6}{\giga\hertz}$ to $\SI{12}{\giga\hertz}$ are shown in Figure~\ref{fig:img2_man_patch}(b) and (c), respectively. Similar to Figure~\ref{fig:img_man_patch}(b), the passive tag on the abdomen of the mannequin is clearly visible, as highlighted by the red circles in Figure~\ref{fig:img2_man_patch}(b). However, because the tags are not resonant within the higher frequency band, they cannot be seen in Figure~\ref{fig:img2_man_patch}(c), despite the increased resolution. By designing tags with different scattering properties, users can select tags that are not visible to systems with a limited frequency range. This enables a customizable approach, allowing scattering intensity and visibility to be adjusted according to user preferences, thereby enhancing both privacy and control over the localization system.

A comprehensive evaluation framework for localization systems is proposed in~\cite{zafari2019survey}, which compares techniques in terms of availability, cost, energy efficiency, scalability, localization accuracy, reception range, and latency. The presented technique demonstrates advantages in terms of availability, since it is based on the concept of passive radar, which does not require complex infrastructure aside from the measurement devices. The method can be deployed conveniently in various application environments, including large spaces such as offices and hospitals, with presence of ubiquitous Tx signals. In terms of cost and energy efficiency, the passive tags are highly economical and do not require a power supply or battery. These features also enhance the scalability of the approach, where the measurement devices may, however, introduce additional costs and consume some energy. Depending on the requirements and available measurement resources, the localization accuracy can be customized over a wide range, up to micro-location accuracy~\cite{zafari2016microlocation}. The passive tags also help to extend the reception range in the considered NF scenarios. While a larger reception range may reduce localization accuracy, this can be compensated by increasing the size of the measurement aperture and the utilized bandwidth, as indicated in~\eqref{eq:resolution}. Regarding latency, despite the highly-efficient implementation of the imaging algorithm, real-time localization remains challenging due to the measurement process. Nonetheless, the approach presents an alternative technical route compared to existing methods, offering enhanced privacy considerations and flexible customization options.
% \section{Privacy Consideration}
% \label{sec:privacy}

\section{Conclusion}
\label{sec:conclusion}
An indoor localization method based on a 3-D holographic passive radar imaging technique has been discussed and illustrated. This localization method utilizes passive tags that rely solely on electromagnetic scattering, and eliminate the need for traditional signal transmission or response mechanisms commonly used in technologies like RFID. The approach allows for precise localization in an intuitive manner within complex indoor geometries, even with limited collected data for imaging. The passive tags are pure electromagnetic scatterers without any integrated circuits or chips, making them easy to adapt and manufacture. Identification of people or objects is also possible through different geometrical properties of the passive tags when imaging resolution is sufficient. Privacy protection mechanisms specific to this technique are also considered. For example, by carrying tags with distinct shapes and frequency behaviors, the users themselves can decide to what level they can be tracked and identified. The method has, thus, the potential to enable versatile indoor localization while maintaining privacy protection capabilities.
% \subsection{Figures}
% For the best publishing quality, we recommend the EPS format for figures. You can also insert figures in doc file or provide figures in PDF, JPG, TIFF, etc. formats. Please note that the original figures' quality is crucial for the final quality of the paper.
% Figs.~\ref{fig:test1} and \ref{fig:test2} show how to insert an encapsulated postscript figure.

% \begin{figure}[h]
%  \centerline{\includegraphics[width=0.56\columnwidth,draft=false]{Figs/fig1.eps}}
%  \caption{This is a sample figure.}
%  \label{fig:test1}
% \end{figure}

%  \begin{figure*}[t]
%  	\begin{center}
%  	\begin{minipage}[t]{0.48\linewidth}
%  		\centering
%  		\begin{overpic}[width=0.55\columnwidth]{Figs/fig2a.eps}
%  			\put(-10,52){\footnotesize{(a)}}
%  		\end{overpic}
%  	\end{minipage}\hspace{-3cm}
%  	\begin{minipage}[t]{0.48\linewidth}
%  		\centering
%  		\begin{overpic}[width=0.55\columnwidth]{Figs/fig2b.eps}
%  			\put(-10,52){\footnotesize{(b)}}
%  		\end{overpic}
%  	\end{minipage}
%   \end{center}
%  	\caption{This is a sample subfigure.}
%  	\label{fig:test2}
%  \end{figure*}

% \subsection{Tables}

% Table~\ref{tab1} shows how to display a table.

% \begin{table}[h]
% \renewcommand{\arraystretch}{1.3}
% \caption{A sample table.}
% \begin{center}
% \begin{tabular}{|c|c|c|}
% \hline
% col 1's label & col 2's label & col 3's label \\ 
% \hline
% 1 & 2 & 3 \\ 
% \hline
% 130 &0.5 & 1.159 \\
% \hline
% \end{tabular}
% \end{center}
% \label{tab1}
% \end{table}

% \section{Citations}

% The references are cited as \cite{2Lastname,3Lastname}.

\section{Acknowledgement}

Funded by the European Union. Views and opinions expressed are, however, those of the author(s) only and do not necessarily reflect those of the European Union or European Innovation Council and SMEs Executive Agency (EISMEA). Neither the European Union nor the granting authority can be held responsible for them. Grant Agreement No: 101099491.

% \appendixx{This is the First Appendix}

% \subsection{Sub-Appendix}

% \begin{equation}
%  E=mc^2
% \end{equation}

% \appendixx{This is the Second Appendix}

% \begin{equation}
%  F=ma
% \end{equation}

 \nocite{*}
 \bibliographystyle{jpier}
 \bibliography{refs}

% \end{multicols}
\end{document}